\documentclass[conference]{IEEEtran}
 
\usepackage{cite}

\usepackage{xcolor}
\usepackage{tikz}
\usetikzlibrary{calc,positioning}

\usepackage{graphicx}
\usepackage{amsmath}
\usepackage{url}

\usepackage[]{fancyhdr} %

\ifCLASSOPTIONcompsoc
 \usepackage[caption=false,font=normalsize,labelfont=sf,textfont=sf]{subfig}
\else
 \usepackage[caption=false,font=footnotesize]{subfig}
\fi
\IEEEoverridecommandlockouts
\begin{document}

%
\title{Statistical Modeling and Forecasting of \\Automatic Generation Control Signals}

\author{
\IEEEauthorblockN{Sarnaduti Brahma, Hamid R. Ossareh, and Mads R. Almassalkhi}
\IEEEauthorblockA{Department of Electrical and Biomedical Engineering \\
The University of Vermont (UVM)\\
Burlington, VT 05405, USA\\
\{sbrahma, hossareh, malmassa\}@uvm.edu}
}


%



\lhead{ACCEPTED FOR PRESENTATION IN 11TH BULK POWER SYSTEMS DYNAMICS AND CONTROL SYMPOSIUM (IREP 2022), JULY 25-30, 2022, BANFF, CANADA}


\maketitle
\thispagestyle{fancy}
\pagestyle{fancy}


\begin{abstract}
The performance of frequency regulating units for automatic generation control (AGC) of power systems depends on their ability to track the AGC signal accurately. In addition, representative models and advanced analysis and analytics can yield forecasts of the AGC signal that aids in controller design. In this paper, time-series analyses are conducted on an AGC signal, specifically the PJM Reg-D, and using the results, a statistical model is derived that fairly accurately captures its second moments and saturated nature, as well as a time-series-based predictive model to provide forecasts. \textcolor{black}{As an application}, the predictive model is used in a model predictive control framework to ensure optimal tracking performance of a down ramp-limited \textcolor{black}{distributed energy resource} coordination scheme. The results provide valuable insight into the properties of the AGC signal and indicate the effectiveness of these models in replicating its behavior.
\end{abstract}



%
\IEEEpeerreviewmaketitle

\section{Introduction}
In electric power systems, correcting the mismatch between demand and supply is crucial for reliable operation \cite{Kundur}. This is usually achieved through regulation services. Recently developed coordination and control schemes allow distributed energy resources (DERs), such as air conditioners and electric water heaters, to provide regulation services such as frequency regulation using automatic generation control (AGC) \cite{Almassalkhi:2018IMA,Callaway:2011wq,energies,optres}. In such a coordinating scheme, an aggregate of regulating resources tracks an AGC signal, resulting in the frequency being maintained at the required value. Moreover, due to the increasing penetration of renewable sources of energy like solar/wind, there is added variability (due to the fluctuating nature of the sources) and uncertainty (due to lack of accurate forecasts) in the demand \cite{balancing}. When the \textcolor{black}{generation from variable renewables} is under-predicted, more generation is scheduled than necessary, leading to increased costs. When it is over-predicted, less generation is scheduled, requiring more expensive quick-start power generators and/or load shedding. Hence, there is a need for balancing authorities like the Pennsylvania-New Jersey-Maryland interconnection (PJM) ~\cite{pjmmanual} that coordinate these resources effectively to maintain the balance between demand and supply.

By considering the characteristics of the specific DER coordination scheme, controllers can be designed to guarantee high tracking performance of the regulation resources and, thus, ensure optimal and reliable operation. As a case in point, the authors are involved in a flexible load coordination scheme called Packetized Energy Management (PEM) \cite{Almassalkhi:2018IMA,almassalkhi2017packetized}, which is a demand dispatch scheme where loads individually request and can be granted uninterruptible access to the grid for a pre-specified time interval called the packet length. Such a bottom-up coordination method centers on preserving the end-user quality of service (QoS). However, this results in a down ramp-limited response because the loads only consume power from the grid and do not transition from on to off until their pre-specified interval is completed. By taking into account the down ramp-limited nature of PEM, a predictive control scheme can be designed, which can precompensate the AGC signal, using knowledge of its future values, to improve the tracking performance of PEM. \textcolor{black}{The tracking performance of up-ramp limited thermal generators can be likewise improved in this manner.} To maximize the performance of PEM and other such demand dispatch schemes, however, requires modeling and forecasting of the AGC signal.

Controllers for frequency regulating units can be more effectively designed if there is knowledge of the statistical properties of the specific AGC signal being tracked. Of commonly available AGC regulation signals, the Reg-D signal, provided by the regulatory authority, PJM, which is part of the Eastern Interconnection in the United States, is an ``energy-neutral'' regulation AGC signal, typically dispatched every two seconds.  Compared to the Reg-A, which is another, slower, PJM regulation signal that is sent to traditional resources and meant to recover larger, longer fluctuations in system conditions, the Reg-D is a fast, dynamic signal that is sent to dynamic resources. Its hourly average tends toward zero (i.e., it is \textit{energy-neutral}), but it requires resources to respond rapidly~\cite{pjmmanual}. Reg-D is normalized between $-1$ and $1$, with $-1$ and $1$ representing minimum and maximum power capacity (MW) bid into the frequency regulation market by an aggregator, respectively. To the best knowledge of the authors, there is currently no work in the literature that provides a detailed statistical analysis of AGC signals, such as PJM Reg-D, intending to derive accurate models and forecasts, which are essential for effective controller design. \textcolor{black}{The paper \cite{agcmod} provides brief  analyses on a specific AGC signal from the Bonneville Power Administration (unlike on the commonly available AGC signal PJM Reg-D attempted here), specifically regarding its statistical distribution and change in energy content across hours. However, unlike this paper, \cite{agcmod} does not conduct other important statistical analyses on the AGC signal or provide a statistical model. The paper \cite{agcmod} also provides ARMA forecasting models to predict the \textit{hourly energy content} of the AGC signal (unlike the AGC signal itself that is predicted here), and describes a method to predict the state-of-charge of an energy storage resource based on forecasts of the AGC hourly energy content. However, it does not provide simulations regarding the effectiveness of the forecasting model on the practical application (of predicting the state-of-charge).} 

This paper fills the above gap. Specifically, we investigate the statistical properties of a widely known AGC signal, PJM Reg-D, intending to develop two models: a statistical model that fairly accurately captures the second moments of the Reg-D signal and its saturated nature, and a time-series-based forecasting model. A statistical model for the AGC signal enables the model-based design of controllers by providing accurate representations of its variability and/or saturated nature \cite{qlc,chen1998multivariate}, whereas a time series based forecasting model can be used to predict the future AGC signal (either its value or direction) and make decisions on allocating resources effectively based on that prediction, including designing model predictive controllers. For example, when there are steep ramp-ups followed immediately by ramp-downs in the AGC signal, \textcolor{black}{tracking them optimally can maximize performance score, which increases revenue under pay-for-performance schemes.} If it can be predicted that the AGC signal will ramp up and down quickly in the future, then resources can be optimally utilized. 
The analyses in this paper are conducted on a year-long historical data of the signal from July 2018 to June 2019 with a 2~s resolution obtained from PJM \cite{dataminer}.

First, the statistical distribution of Reg-D is investigated. 
Second, the variability and stationarity of Reg-D are investigated. The mean-variance of AGC across minutes, hours, days, and months are computed, as well as the running mean and variance. 
Third, the statistics of saturation of Reg-D (i.e., values $-1$ and $1$), known as ``pegging", are evaluated. It is found that the amount of pegging is directly related to the variance of Reg-D. Fourth, the power spectral density of Reg-D is computed for different months.

Fifth, using the information on mean, variance, stationarity, and bandwidth of Reg-D, a stochastic model is constructed that consists of zero-mean stationary white Gaussian noise passed through a coloring filter of appropriate bandwidth, and the output scaled by the standard deviation of Reg-D and saturated between $-1$ and $1$. It is found that this statistical model fairly accurately captures the second moments of Reg-D (with $<3.5$\% error) and its saturated nature (with $<2.5$\% error).

Sixth, using the autocorrelation and partial autocorrelation functions of Reg-D,
an autoregressive moving average (ARMA) model for predicting Reg-D is designed. It is found that an AR(3) model can provide a directionally salient prediction, i.e., the slope of Reg-D and the ARMA forecast are highly correlated (correlation coefficient $>0.5$) for up to 30 s.

Finally, to further improve forecasts, the cross-correlation of AGC with frequency is evaluated using the additional historical data obtained from phasor measurement units (PMUs) in PJM's territory. Specifically, a vector autoregressive moving average (VARMA) model is constructed using both Reg-D and frequency data. It is found that a VAR(3) model can provide a significant improvement in the forecasts (by about 3.5\% over a horizon of 1 min) compared to the corresponding AR(3) model. To illustrate the effectiveness of the forecasts in a practical application, the VAR(3) forecasts are also employed in a model predictive controller (MPC) described in \cite{brahma2021optimal}. \textcolor{black}{The MPC designed in \cite{brahma2021optimal} uses AR(3) predictions of the power output of a DER-coordination scheme, Packetized Energy Management (PEM), to predict its down-ramp-limited nature and pre-compensate the AGC input to improve its tracking performance, compared to the case with no precompensator.} In this paper, it is shown that the tracking performance of this down ramp-limited demand dispatch scheme with MPC can be improved by around 1\% compared to AR(3) forecasts.
The original contributions of this paper are, thus, as follows:
\begin{itemize}
    \item Statistical analysis is systematically conducted on the AGC regulation signal, PJM Reg-D.
    \item Using the results of the analysis, a linear stochastic model of AGC is derived that is driven by stationary white noise,
    \item Time series-based forecasting models are developed using ARMA and VARMA models that are effective in predicting the future values of the AGC signal.
    \item The VARMA-based forecasting model is applied to a model predictive controller for the Packetized Energy Management scheme from \cite{brahma2021optimal} to indicate the effectiveness of the forecasts and the resulting improvement in tracking performance compared to ARMA forecasts.
\end{itemize}

\section{Statistical Modeling of AGC signal}
\begin{figure}[t]
\centering
\subfloat[Full data]{\includegraphics[width=0.75\columnwidth]{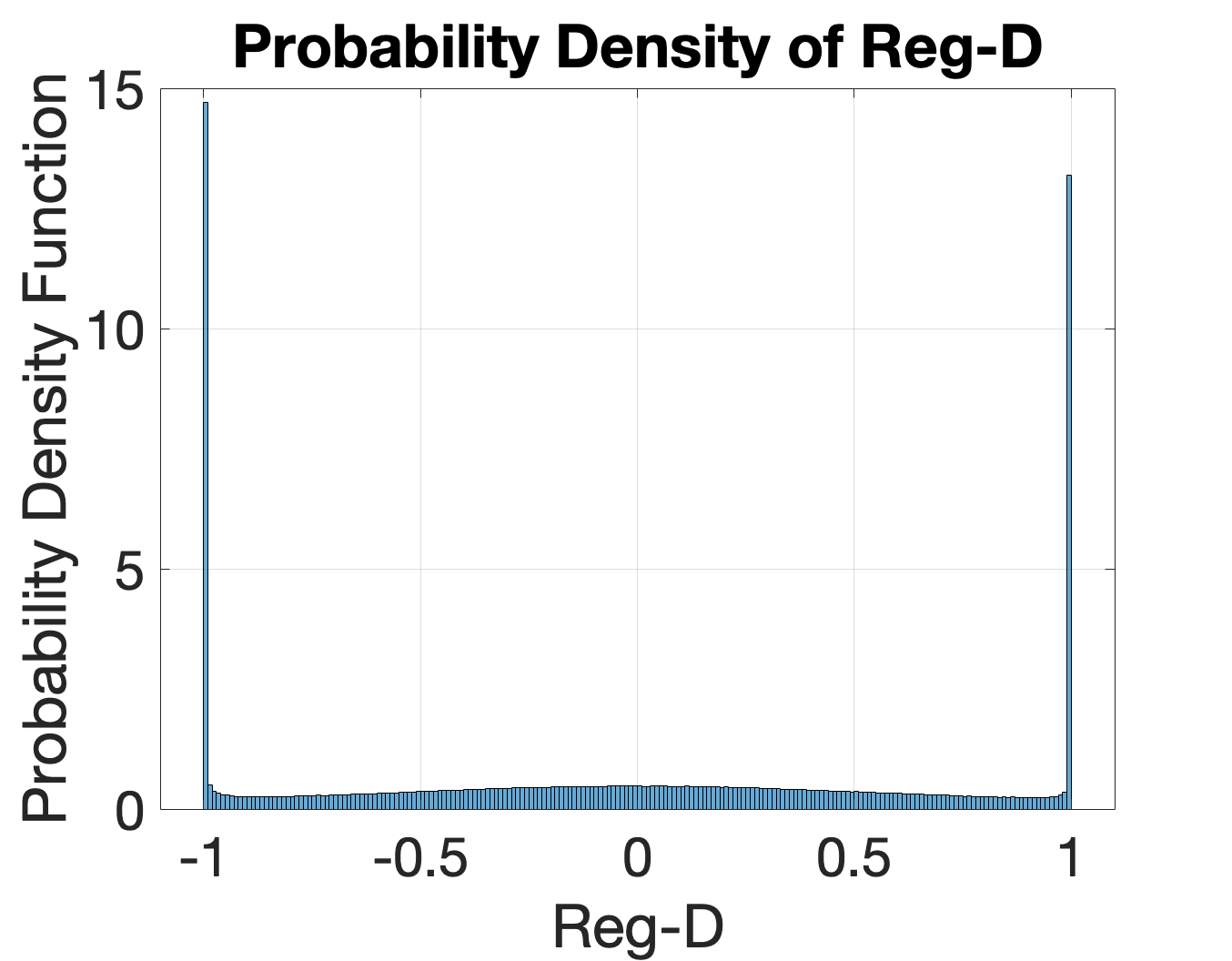}\label{fig:regdfull}}
\hfil
\subfloat[Excluding values near -1 or 1]{\includegraphics[width=0.75\columnwidth]{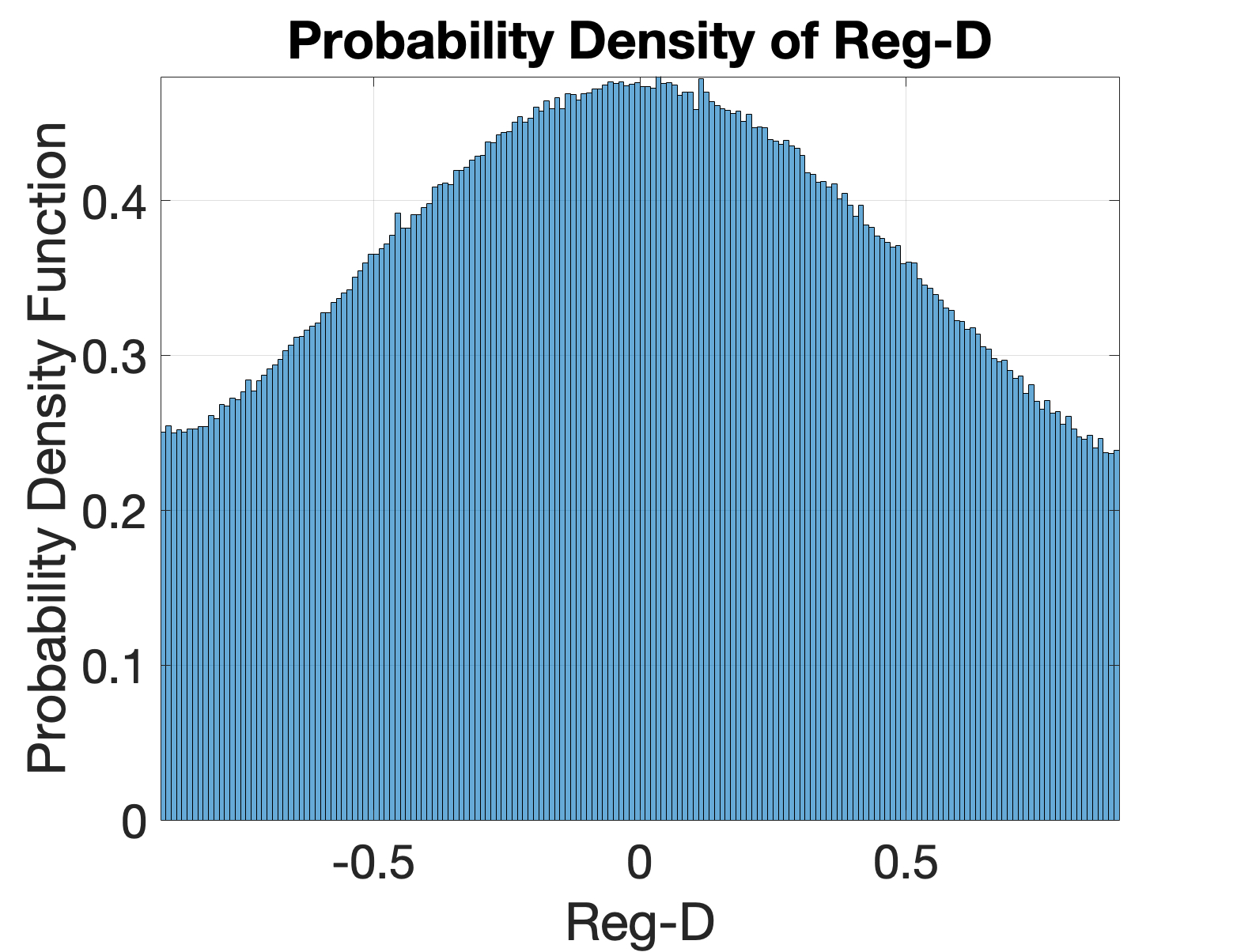}\label{fig:regdzoom}}
\caption{Probability Density of Reg-D}
\label{fig:regdprob}
\end{figure}
This section aims to derive a stochastic model of an AGC signal, specifically the PJM Reg-D, driven by a random noise process. To motivate the appropriate form of the statistical model, statistical analyses are first conducted on the Reg-D signal. \textcolor{black}{Specific analyses on the variability of Reg-D across different time scales were conducted in \cite{brahma2021optimal}, the details of which have been omitted here.}

\subsection{Statistical Distribution}\label{ssec:dist}
First, to obtain an idea of the distribution of the values of  Reg-D, its histogram was plotted (using the probability density normalization) on the data for the entire year from July 2018 to June 2019. It can be seen from Fig. \ref{fig:regdfull} that the signal is mostly saturated at $1$ or $-1$. However, on zooming near the value of zero (Fig. \ref{fig:regdzoom}), it can be seen that the distribution of Reg-D can be approximated to be a zero-mean truncated Gaussian \cite{gubner2006probability}.

\subsection{Wide-sense Stationarity and Ergodicity}\label{ssec:wss}
\begin{figure}
    \centering
    \subfloat[Sample Mean]{\includegraphics[width=0.6\columnwidth]{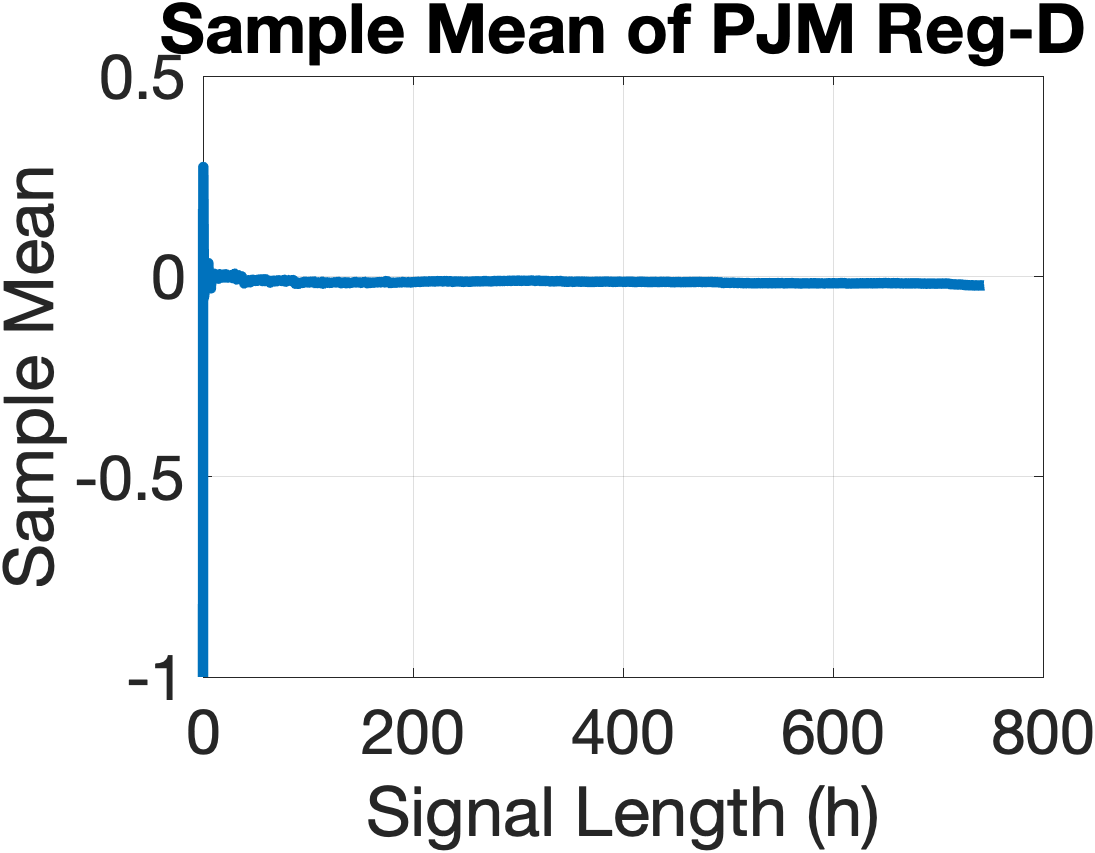}\label{fig:sampmean}}
    \hfil
    \subfloat[Sample Variance]{\includegraphics[width=0.6\columnwidth]{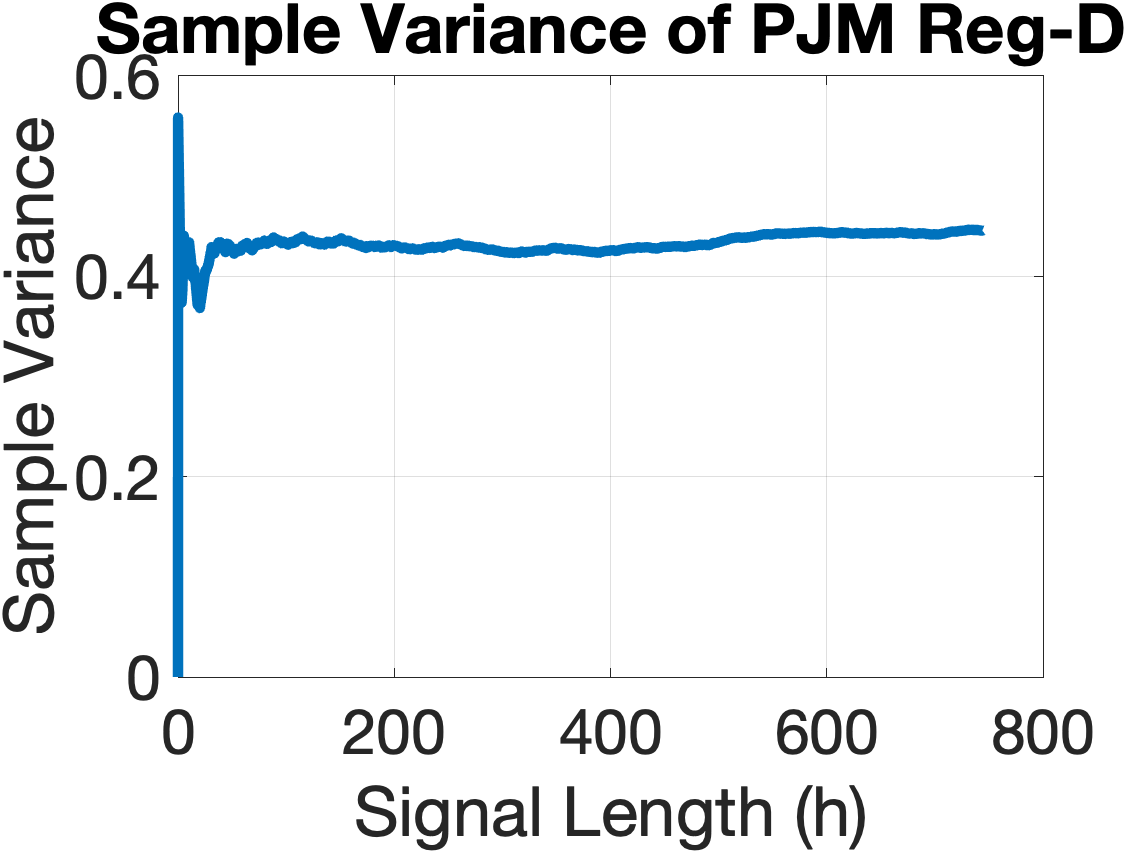}\label{fig:sampvar}}
    \caption{Sample Mean and Variance of PJM Reg-D}
    \label{fig:sampmeanvar}
\end{figure}
    
\begin{figure}
    \centering
    \subfloat[Settling Time of Sample Mean]{\includegraphics[width=0.9\columnwidth]{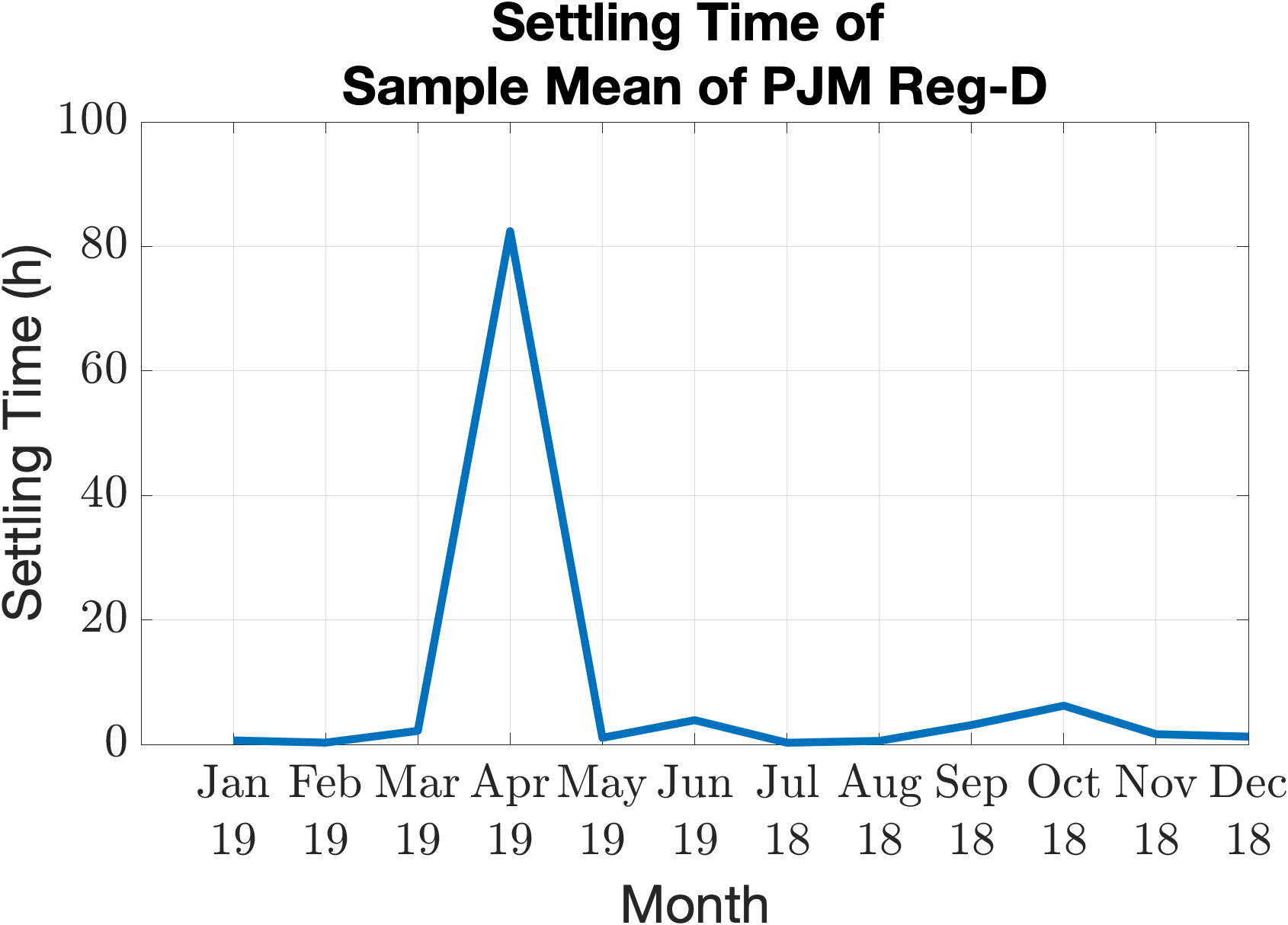}\label{fig:sampmst}}
    \hfil
    \subfloat[Settling Time of Sample Variance]{\includegraphics[width=0.9\columnwidth]{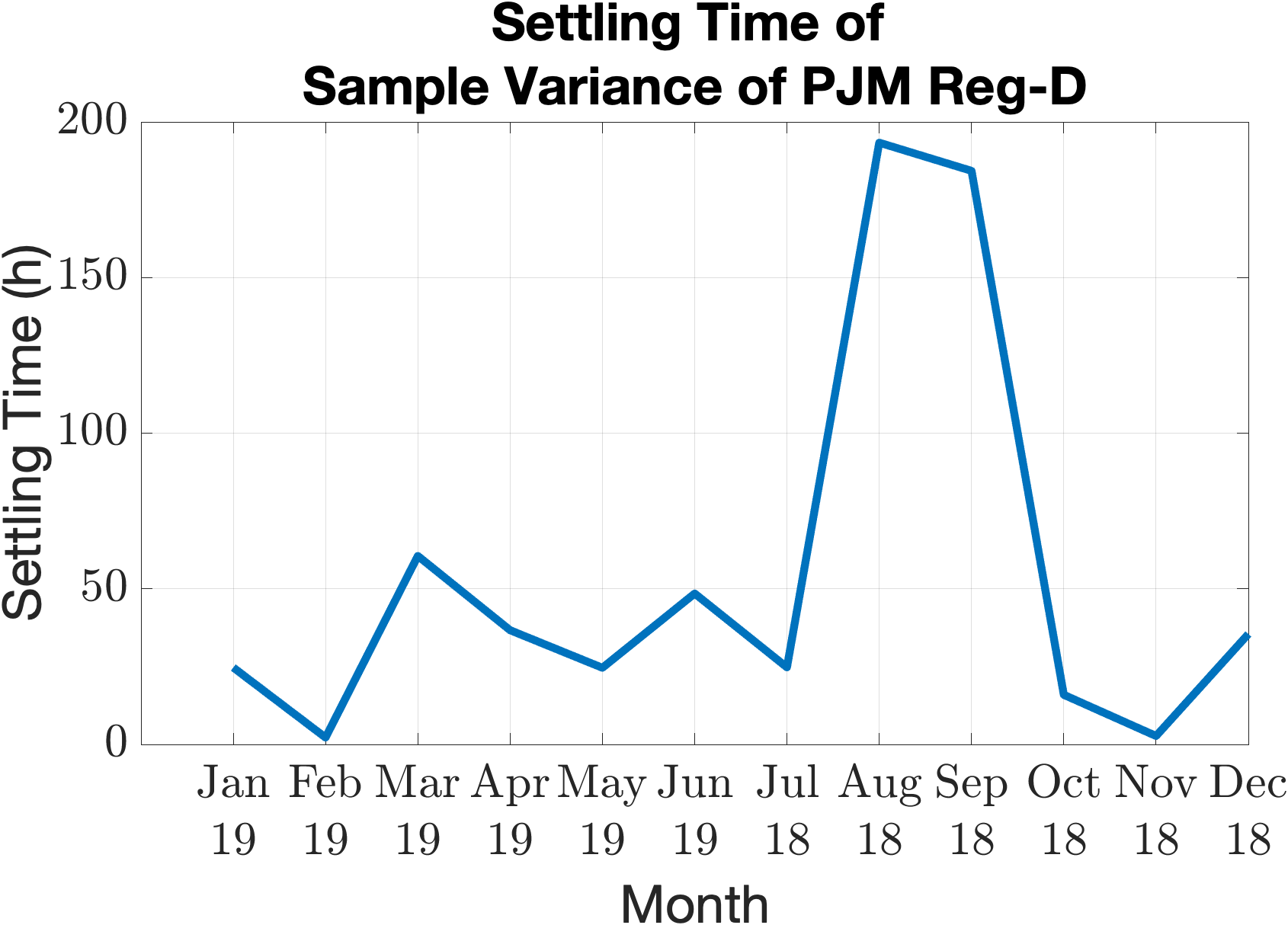}\label{fig:sampvst}}
\caption{Settling Time of Sample Mean and Variance of PJM Reg-D}
\label{regdmeanvarst}
\end{figure}

Next, to determine whether a statistical model for Reg-D can be driven by a \textit{stationary} random process and whether the model will be valid across multiple time ranges, the stationarity of PJM Reg-D was investigated. Specifically, the sample mean and variance of Reg-D are plotted in Figs. \ref{fig:sampmean} and \ref{fig:sampvar}. It can be seen that after about 1 hour, the sample mean settles to a value close to 0, while the sample variance settles after about 25 hours to a value close to 0.42. To quantify this systematically, the settling times (i.e., the time when the signal is within 2\% of its final value) of the sample mean and sample variance of Reg-D are evaluated at different months of the year. The results are shown in Figs. \ref{fig:sampmst} and \ref{fig:sampvst}. \textcolor{black}{It was found that the settling time of the sample mean is, on average, less} than two hours, except for April 2019 (during April 2019, it was observed that there were more peg-up events than peg-down events). \textcolor{black}{Since the mean is close to zero, it indicates that Reg-D is \textit{energy-neutral}. This reduces the likelihood that an electric storage resource would have insufficient energy to respond to Reg-D, thereby reducing its potential
compensation and ability to provide regulation in a future interval \cite{pjmmanual}. The above results indicate that Reg-D is fairly wide-sense stationary (WSS) and ergodic in the mean and variance, which provides confidence that a WSS statistical model can be considered for it that would be valid across multiple time ranges.}

\subsection{Pegging Amount}\label{ssec:pegam}
\begin{figure}[t]
\centering
\subfloat[Minutely]{\includegraphics[width=0.62\columnwidth]{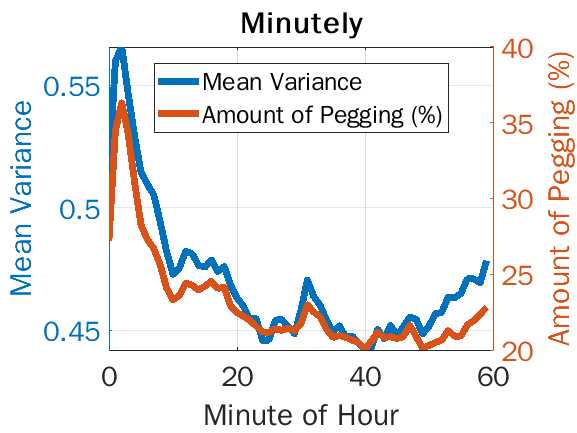}\label{fig:pegmin}}\\
\subfloat[Hourly]{\includegraphics[width=0.62\columnwidth]{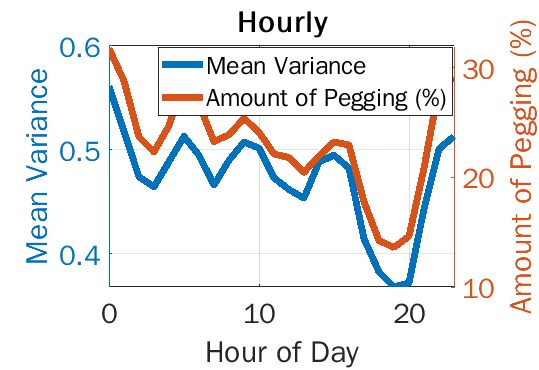}\label{fig:peghour}}
\hfil
\subfloat[Daily]{\includegraphics[width=0.7\columnwidth]{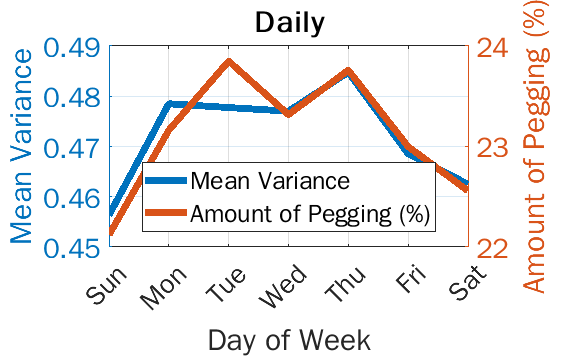}\label{fig:pegday}}\\
\subfloat[Monthly]{\includegraphics[width=0.8\columnwidth]{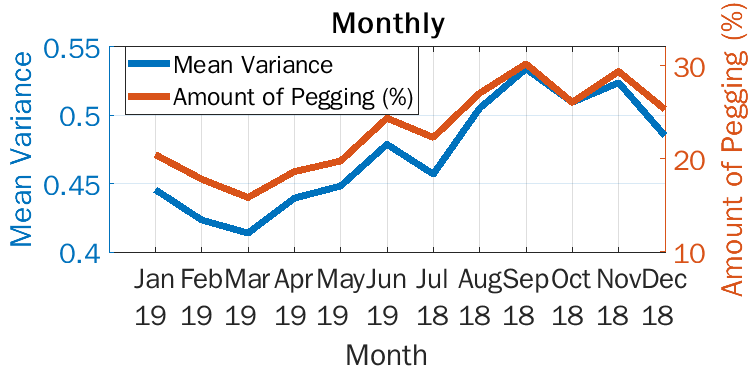}\label{fig:pegmonth}}
\caption{Pegging Amount of PJM Reg-D. For details on the mean variance, please see \cite{brahma2021optimal}.}
\label{regdpeg}
\end{figure}
The Reg-D signal is often saturated between $-1$ and $1$, referred to as ``pegging", which can be due to unexpected and sudden changes in generation or load, unexpected large interchange swings, generation lagging or not following economic dispatch, frequency excursion outside PJM or load forecast error \cite{martini}. The actual AGC signal is usually a scaled and biased version of the Reg-D signal. In this and the following subsections, an analysis of the amount and duration of pegging in the AGC signal are analyzed. This is important since any controller that is specifically designed for non-saturated signals will not perform well when the signal is saturated. In such a case, it will have to be adapted to handle saturation. 

The amount of pegging in the AGC signal is evaluated by finding the percentage of the samples that are saturated over the total number of samples considered in the particular group, i.e., the minute, hour, day, or month. The results are shown in Fig. \ref{regdpeg}. \textcolor{black}{It can be seen that the amount of pegging in the AGC signal is directly related to the variability of Reg-D.}

\subsection{Pegging Duration}\label{ssec:pegdur}
\begin{figure}[t]
\centering
\subfloat[Hourly]{\includegraphics[width=0.7\columnwidth]{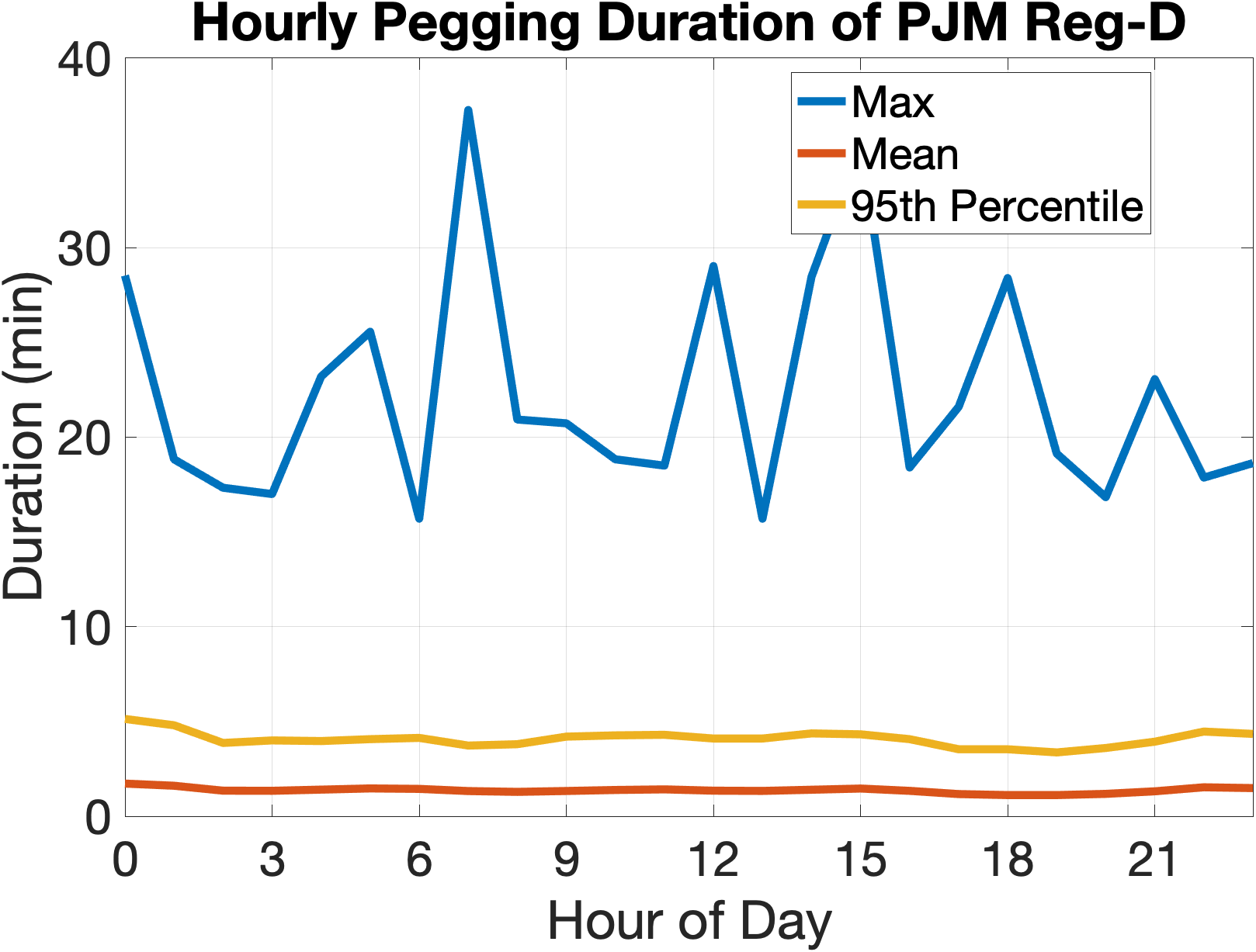}\label{fig:pdurhour}}
\hfil
\subfloat[Daily]{\includegraphics[width=0.7\columnwidth]{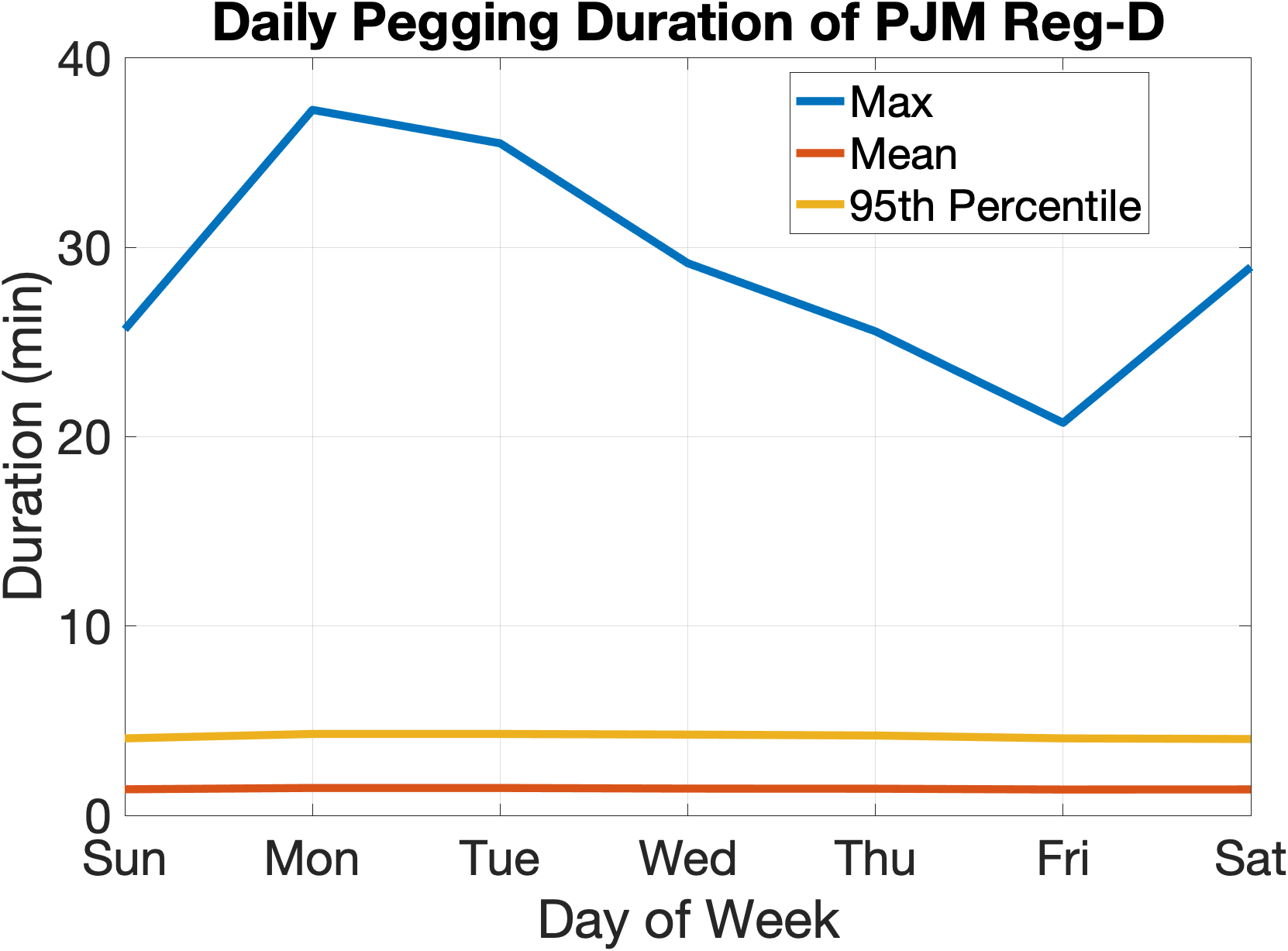}\label{fig:pdurday}}\\
\subfloat[Monthly]{\includegraphics[width=0.8\columnwidth]{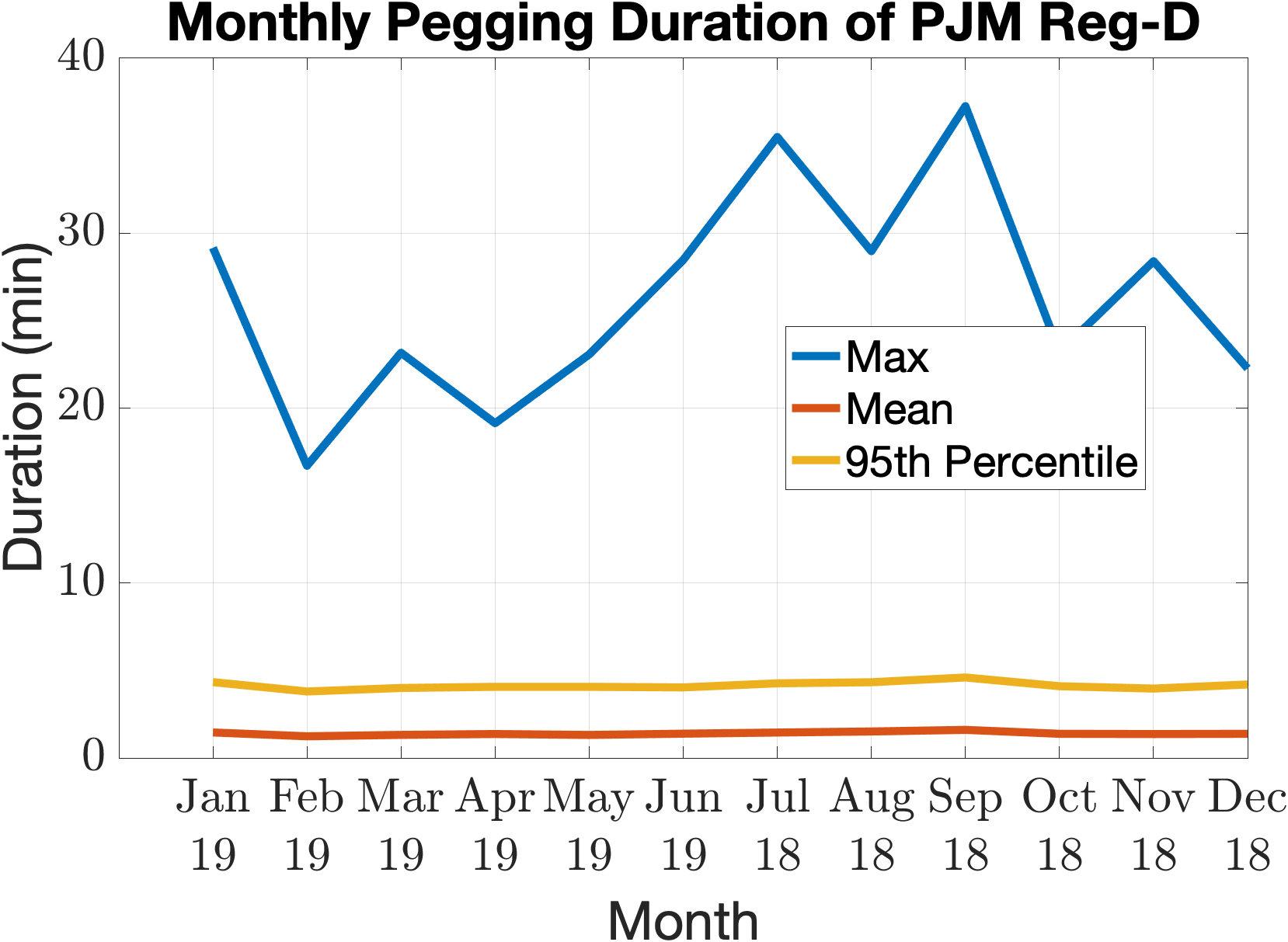}\label{fig:pdurmonth}}
\caption{Pegging Duration of PJM Reg-D}
\label{pegdur}
\end{figure}

Usually, the majority of the pegging events are isolated and of short duration, which may not detrimentally affect the performance of a controller, as compared to long-duration pegging events. To obtain a sense of when and for how long the pegging takes place, apart from the amount of pegging, which was considered above, the continuous pegging duration of the signal was evaluated at different time scales. The results are shown in Fig. \ref{pegdur}. From all the subfigures, it can be seen that the maximum pegging duration can be up to 40 minutes, while the average pegging duration is generally around one minute. Moreover, there is no variability among the different hours, days, and months in the average and the 95th percentile of the pegging duration. However, the maximum pegging duration is seen to be the highest during the start of the week, on Monday, and lowest at the end of the week, on Friday, while the maximum monthly pegging duration is the highest around September and the lowest around February. There is no general trend observed in the maximum hourly pegging duration.

\subsection{Power Spectral Density}\label{ssec:psd}
\begin{figure}[t]
\centering
\subfloat[Spectrum]{\includegraphics[width=0.7\columnwidth]{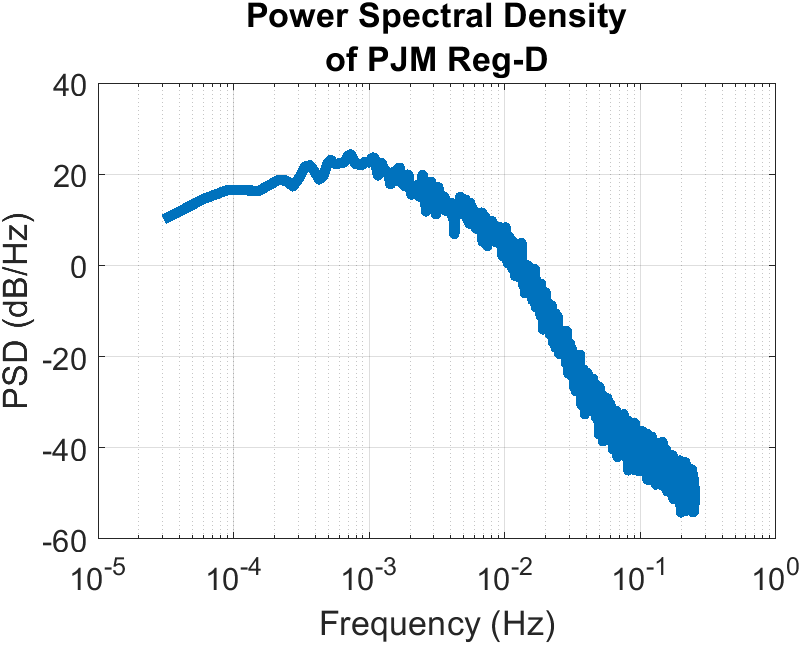}\label{fig:spec}}
\hfil
\subfloat[Bandwidth]{\includegraphics[width=1.05\columnwidth]{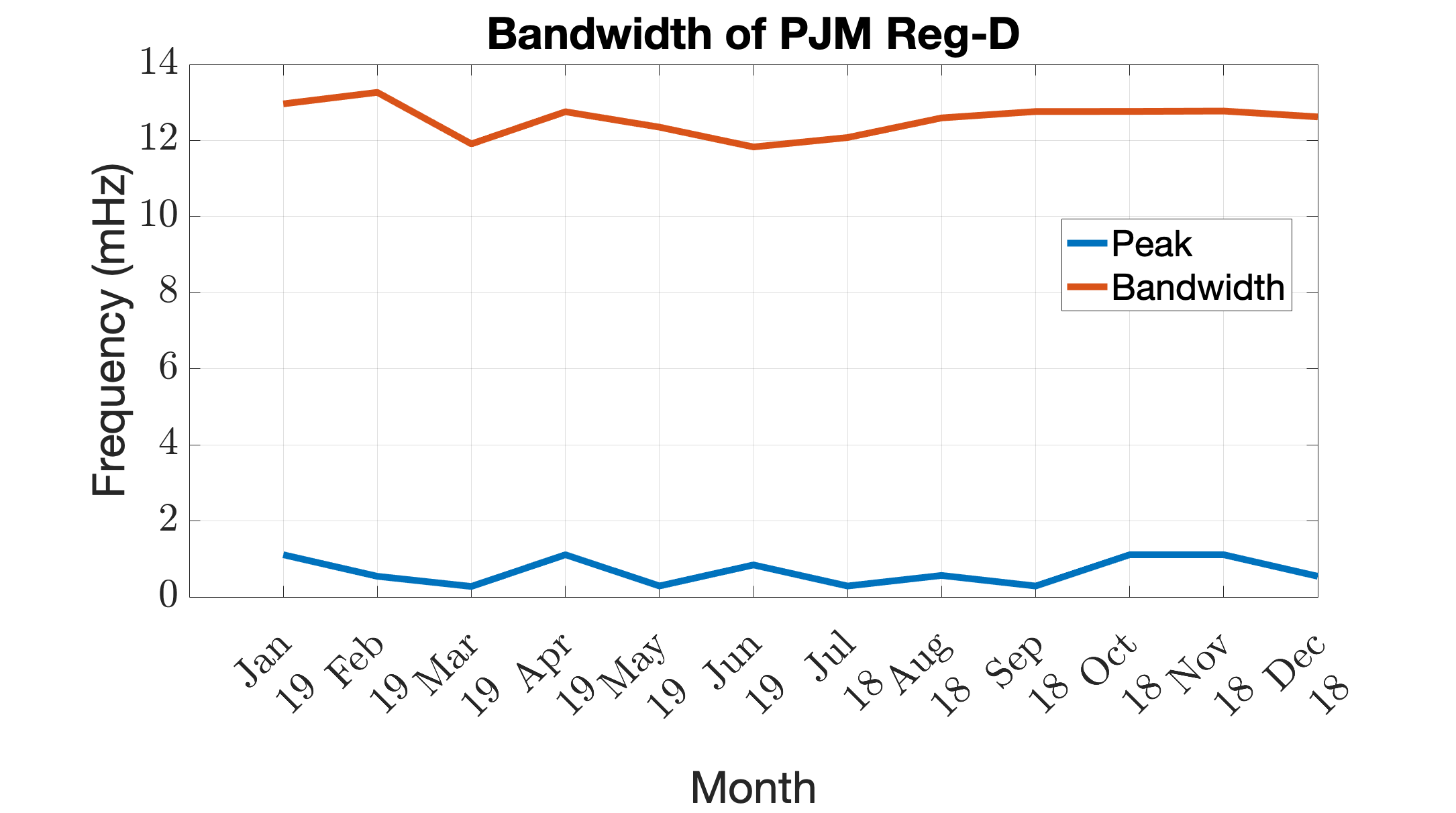}\label{fig:bandwidth}}
\caption{PJM Reg-D Statistics}
\label{forecast}
\end{figure}
Next, the power spectral density (PSD) of Reg-D is analyzed, which provides information about the bandwidth and structure of the filter to be used in the statistical model. The PSD of Reg-D is shown in Fig. \ref{fig:spec}, obtained using the Welch estimate, involving Hanning windows with 50\% overlap \cite{welch}. It can be seen that the signal has a lowpass nature, possibly a slight bandpass nature. The peak of the PSD is at around 0.7 mHz. The 3-dB bandwidth, i.e., the frequency of Reg-D at 3 dB less than the peak PSD of Reg-D, is around 12 mHz. Fig. \ref{fig:bandwidth} shows the peak and bandwidth frequencies across different months of the year 2018-19. It can be seen that both the peak and bandwidth frequencies are relatively constant across the year, which provides evidence that a filter model representing Reg-D can be used across a wide range of periods.

\subsection{Statistical Modeling of AGC}
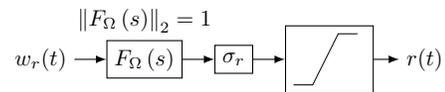
\begin{figure}
    \centering
    \begin{tikzpicture}[>=latex,scale=0.85,every node/.style={transform shape}, every text node part/.style={align=center}]
	\node (wr) at (0,0){$w_r(t)$};
	\node[rectangle,draw] (rf) [right=0.5 of wr,label={above:$\lVert {F_{\Omega}\left(s\right)\rVert}_2=1$}]{$F_{\Omega}\left(s\right)$};
	\node[rectangle,draw] (sr) [right=0.5 of rf]{$\sigma_r$};
	\node[rectangle,draw] (sl) [right=0.5 of sr]{
	\begin{tikzpicture}
	    \draw (0,0)--(0.3,0)--(0.7,0.8)--(1,0.8);
	\end{tikzpicture}
	};
	\draw [->] (wr)--(rf);
	\draw [->] (rf)--(sr);
	\draw [->] (sr)--(sl);
    \draw [->] (sl)--($(sl.east)+(0.4,0)$);
    \node (wr) at ($(sl.east)+(0.8,0)$){$r(t)$};
	\end{tikzpicture}
    \caption{Statistical Model of PJM Reg-D}
    \label{fig:agcmod}
\end{figure}

\begin{figure}[t]
\centering
\subfloat[Time Series]{\includegraphics[width=0.6\columnwidth]{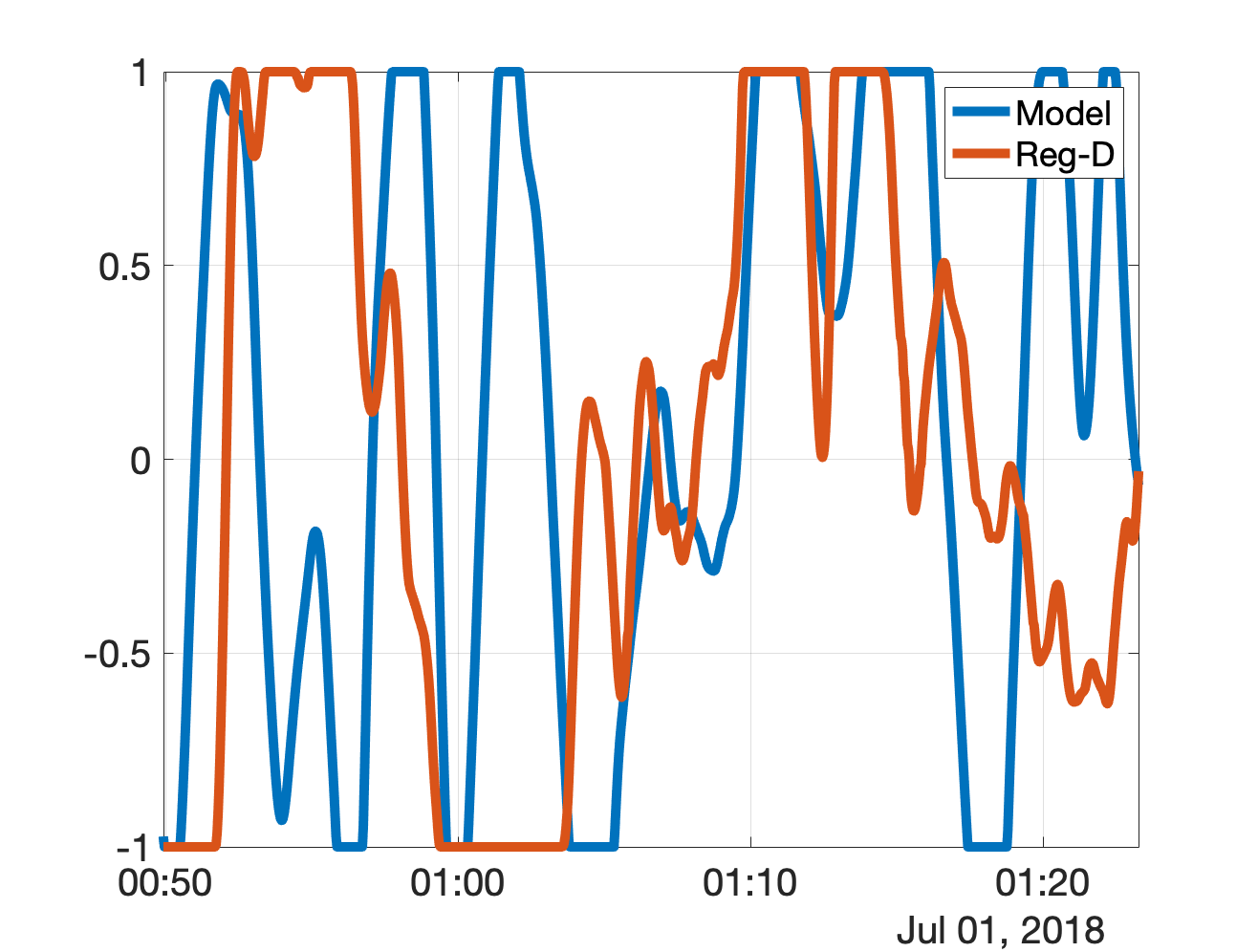}\label{fig:tmregd}}\hfil
\subfloat[\% Error in Standard Deviation]{\includegraphics[width=0.6\columnwidth]{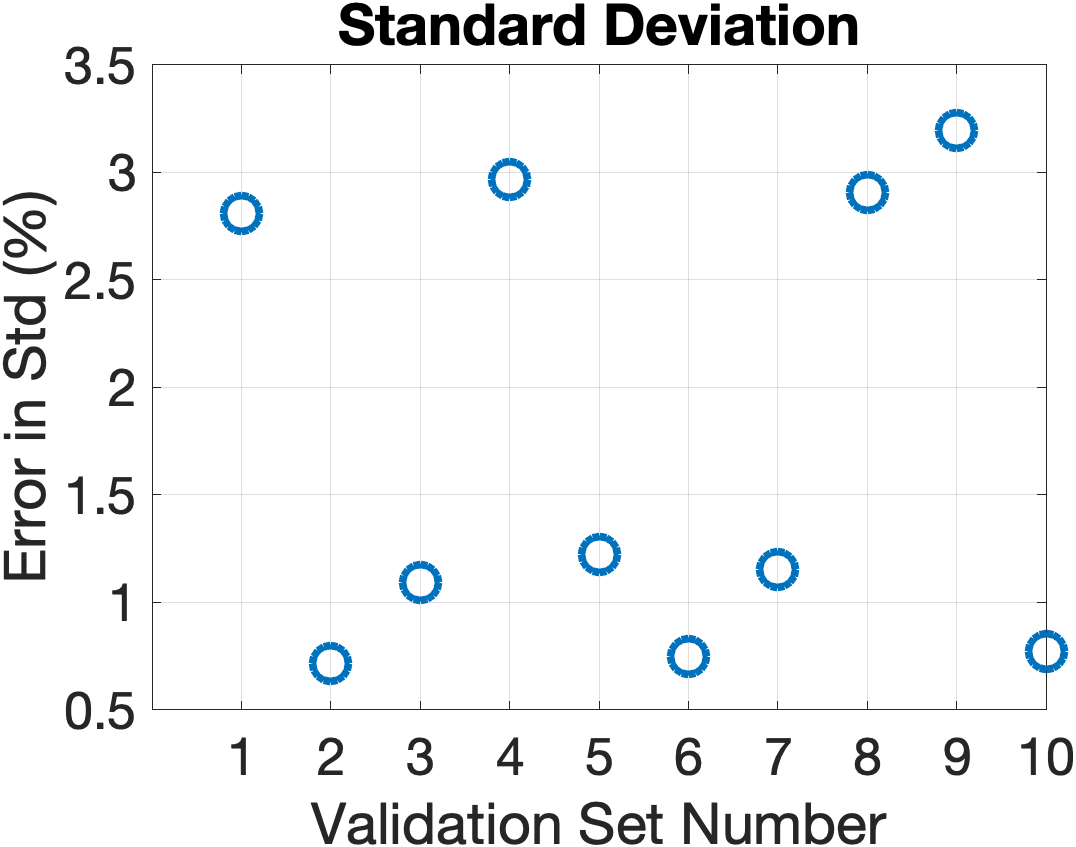}\label{fig:stderr}}\\
\subfloat[Absolute Error in Mean]{\includegraphics[width=0.6\columnwidth]{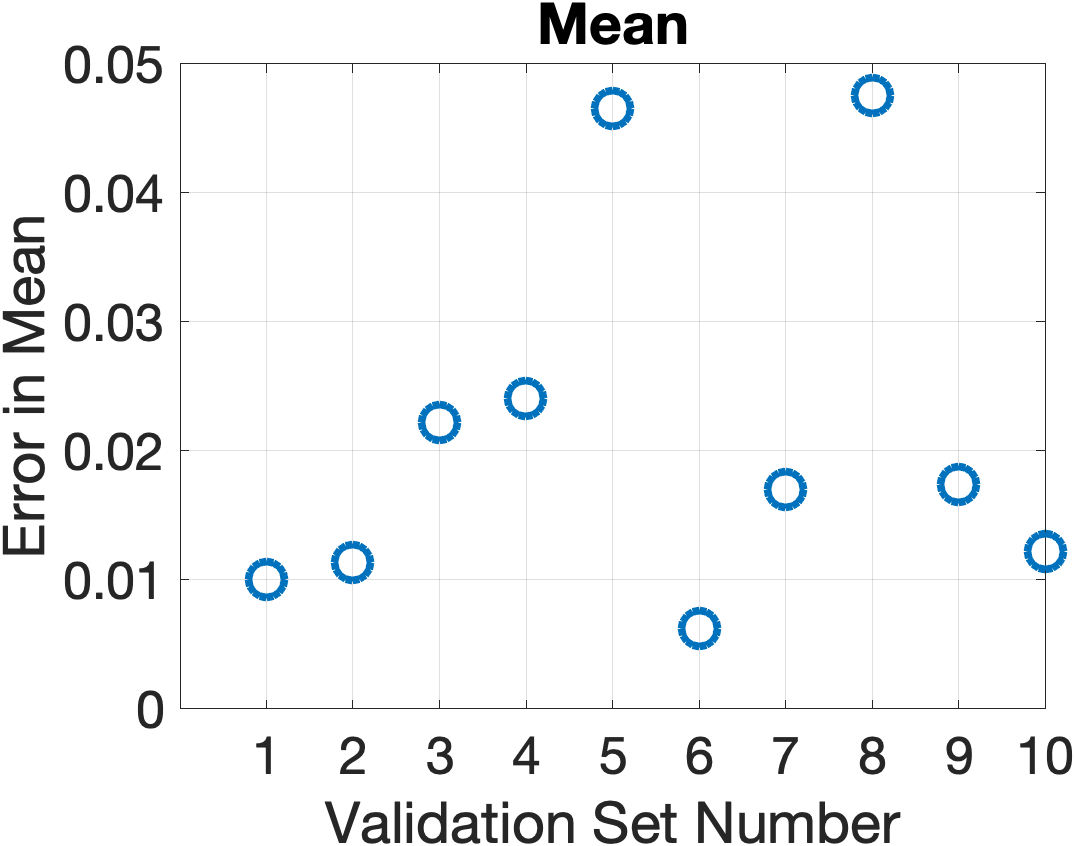}\label{fig:mnerr}}\\
\subfloat[\% Error in Pegging]{\includegraphics[width=0.6\columnwidth]{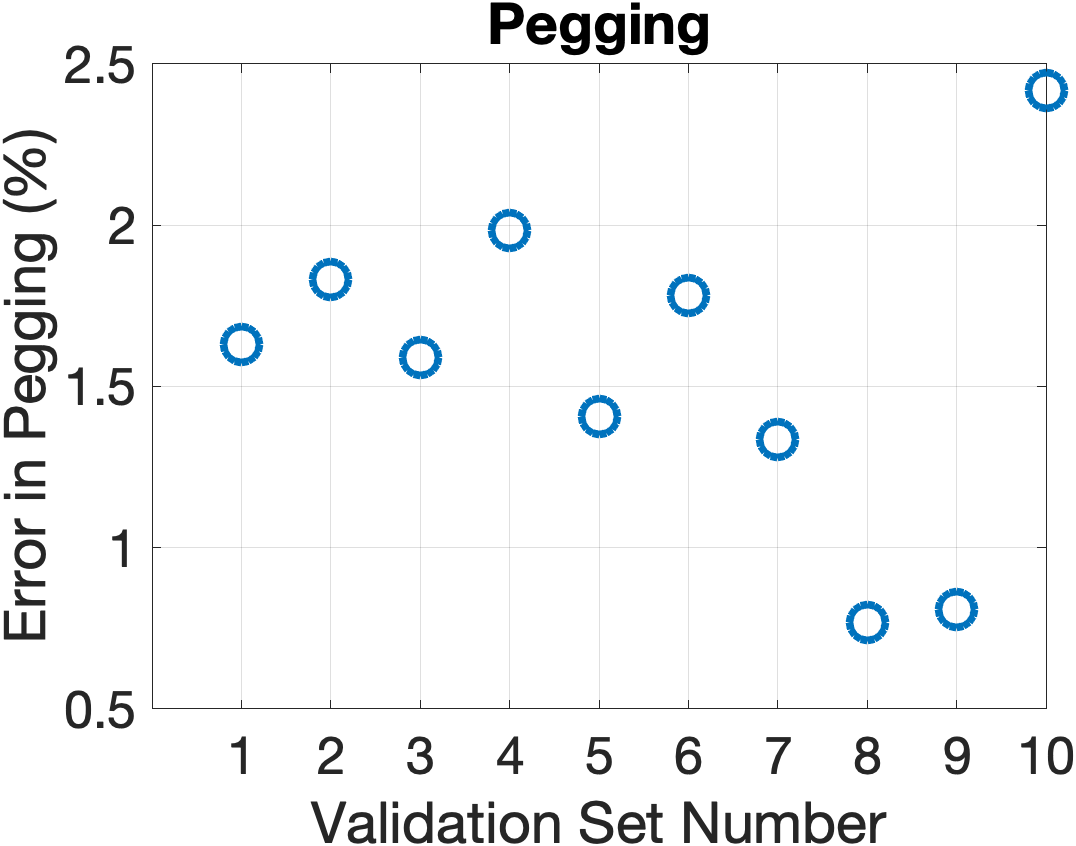}\label{fig:pegerr}}
\caption{Effectiveness of Statistical Model}
\label{modagc}
\end{figure}
\subsubsection{Model} Using the above information, a statistical model of AGC can be derived. The proposed structure of the model is shown in Fig. \ref{fig:agcmod}. It consists of a standard stationary Gaussian white noise process $w_r(t)$ passed through a coloring low pass filter, for example a Butterworth filter, $F_{\Omega}\left(s\right)$ with unit $\mathcal{H}_2$-norm, the output scaled by a constant, $\sigma_r$, and finally the result is saturated between -1 and 1. The rationale for choosing the above elements is explained as follows. Since Reg-D is found to be fairly wide-sense stationary (Section \ref{ssec:wss}), and the values of Reg-D are found to fairly obey a Gaussian distribution for values close to zero (Section \ref{ssec:dist}), the model is driven by a standard Gaussian stationary white noise. The low pass filter is chosen since the PSD of Reg-D (Section \ref{ssec:psd}) informs that it has a low pass nature (which is close in shape to a Butterworth filter in the manner of its roll-off at high frequencies). The bandwidth of the filter can be obtained from the plot of the PSD of Reg-D. The $\mathcal{H}_2$-norm ensures that the variability of the output of the filter is unity, and that it can be scaled to a desired value of variance through the gain, $\sigma_r$, that would be representative of the AGC signal. Finally, the saturation models the pegged nature of Reg-D (Section \ref{ssec:pegam}).

\subsubsection{Validation}
To validate the statistical model of Reg-D, a 3rd order Butterworth filter, with the transfer function,
\begin{equation*}
    F_{\Omega}\left(s\right)=\frac{\sqrt{\frac{3}{\omega_{n}}}}{\left(\frac{s}{\omega_{n}}\right)^{3}+2\left(\frac{s}{\omega_{n}}\right)^{2}+2\left(\frac{s}{\omega_{n}}\right)+1}
\end{equation*}
was chosen to filter the standard stationary white Gaussian noise, where $\Omega_n$ is the bandwidth, taken to be around 5 mHz. A third-order filter is chosen as opposed to, \textcolor{black}{for example}, a first-order filter, since a higher-order filter \textcolor{black}{was found to }result in a smoother signal \textcolor{black}{that is} similar to Reg-D, \textcolor{black}{resulting in a higher fidelity of the statistical model}. Moreover, the filter bandwidth of 5 mHz, which is close to the 3-dB bandwidth obtained from the power spectral density plot, was found to provide a similar rate of fluctuations as in Reg-D (Fig. \ref{fig:tmregd}). The constant, $\sigma_r$, was taken to be 1.25 times the standard deviation of Reg-D. The factor 1.25 was chosen since the AGC signal is saturated and saturation accounts for about 25\% of the signal, as can be seen from Fig. \ref{regdpeg}. Ten 100,000 sample snippets of Reg-D were considered for validation (about 55.5 h each, by which time Reg-D attains fairly wide-sense stationarity, as seen from Fig. \ref{fig:sampmean}-\ref{fig:sampvar}), and the error between the model and the AGC signals in the standard deviation and mean are reported in Figs. \ref{fig:stderr} and \ref{fig:mnerr} for all the validation sets. It can be seen that the error in standard deviation is less than 3.5\% and that in mean is also very small - less than 0.05 in absolute value. Moreover, the error in pegging amount is also small, less than 2.5\%. This indicates that the derived statistical model is accurate both in determining the second moments of the AGC signal, Reg-D, and also its pegged nature.

\section{Forecasting of AGC Signal}
In this section, a forecasting model for an AGC signal is developed. First, an autoregressive moving average (ARMA) model is developed and its effectiveness evaluated. \textcolor{black}{While in \cite{brahma2021optimal}, an ARMA model was briefly described, its effectiveness across different lead times and robustness to coefficients were not explored. A contribution of this section is to provide those analyses and also to use them as a baseline to show improvements with multivariate vector autoregressive moving average (VARMA) forecasts.} Second, using historical frequency data, the cross-correlation between historical AGC and the power grid frequency is evaluated, and a VARMA model is developed. It is shown that the VARMA model can improve prediction performance compared to an ARMA model.

\subsection{ARMA Modeling}
\subsubsection{Model}
To forecast an AGC signal, the following ARMA model can be used \cite{makridakis1997arma}:
\begin{align}
    r[k]={}&\mu+\phi_{1}r[k-1]+\ldots+\phi_{p}r\left[k-g\right]\nonumber\\
     &+a\left[k\right]-\theta_{1}a\left[k-1\right]-\ldots-\theta_{q}a\left[k-h\right]
\end{align}
where $\phi_{i}$ are the autoregressive components, $\theta_i$ the moving average components and $\mu$ is the main level of the process, and $a[k]$ is a stationary zero-mean random Gaussian innovation. $(g,h)$ determines the order of the ARMA model. Since the Reg-D signal is saturated between -1 and 1, the output $r[k]$ of the ARMA model is also saturated between -1 and 1.
The process of determining the order $(g,h)$ of the ARMA model using its autocorrelation and partial correlation functions has been described in Section IIIC of \cite{brahma2021optimal} and is omitted here.


\subsubsection{Forecast Accuracy}\label{ssec:forarmaacc}
\begin{figure}[t]
\centering
\subfloat[Unsaturated ARMA model]{\includegraphics[width=0.6\columnwidth]{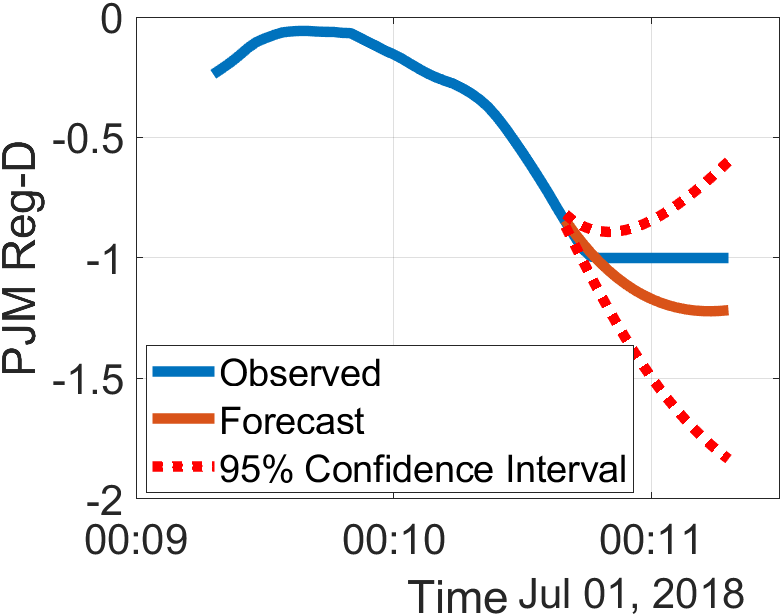}\label{fig:nosat}}
\hfil
\subfloat[Saturated ARMA model]{\includegraphics[width=0.6\columnwidth]{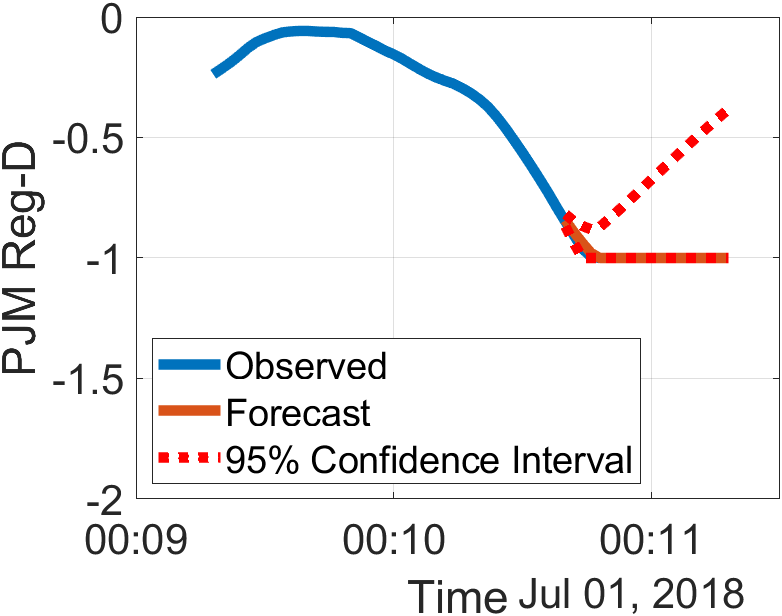}\label{fig:sat}}
\caption{ARMA forecast}
\label{arm:forcast}
\end{figure}

The effectiveness of an AR(3) forecast is shown in Fig. \ref{arm:forcast} for a snapshot of the Reg-D signal in February 2019. Fig. \ref{fig:nosat} shows the forecast if there was no saturation in the ARMA model output. It can be seen that the forecast is outside the acceptable range of -1 and 1, and is hence not valid and leads to a high forecast error. However, if the ARMA model is saturated, the forecast during saturation (Fig. \ref{fig:sat}) is exact compared to the Reg-D signal. Hence, using a saturated ARMA model increases accuracy during pegging by a large amount. From both the figures, however, it can be seen that the forecast is within 95\% confidence of the mean prediction, indicating that it is fairly accurate.

\begin{figure}[t]
    \centering
    \includegraphics[width=0.7\columnwidth]{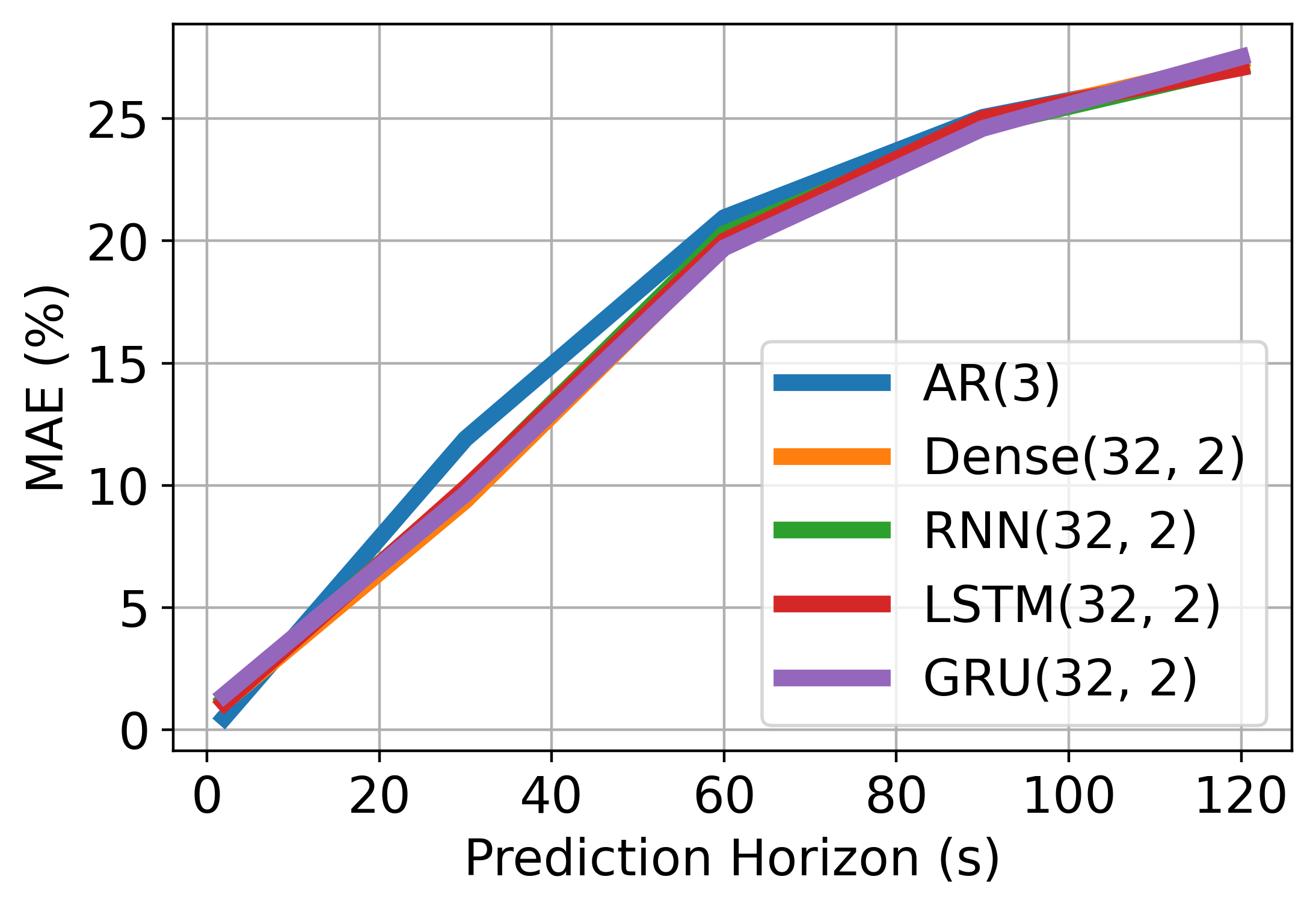}
    \caption{Comparision of AR(3) model to common NN models}
    \label{fig:nn}
\end{figure}

\begin{figure}[t]
\centering
\subfloat[ARMA Forecast accuracy]{\includegraphics[width=0.7\columnwidth]{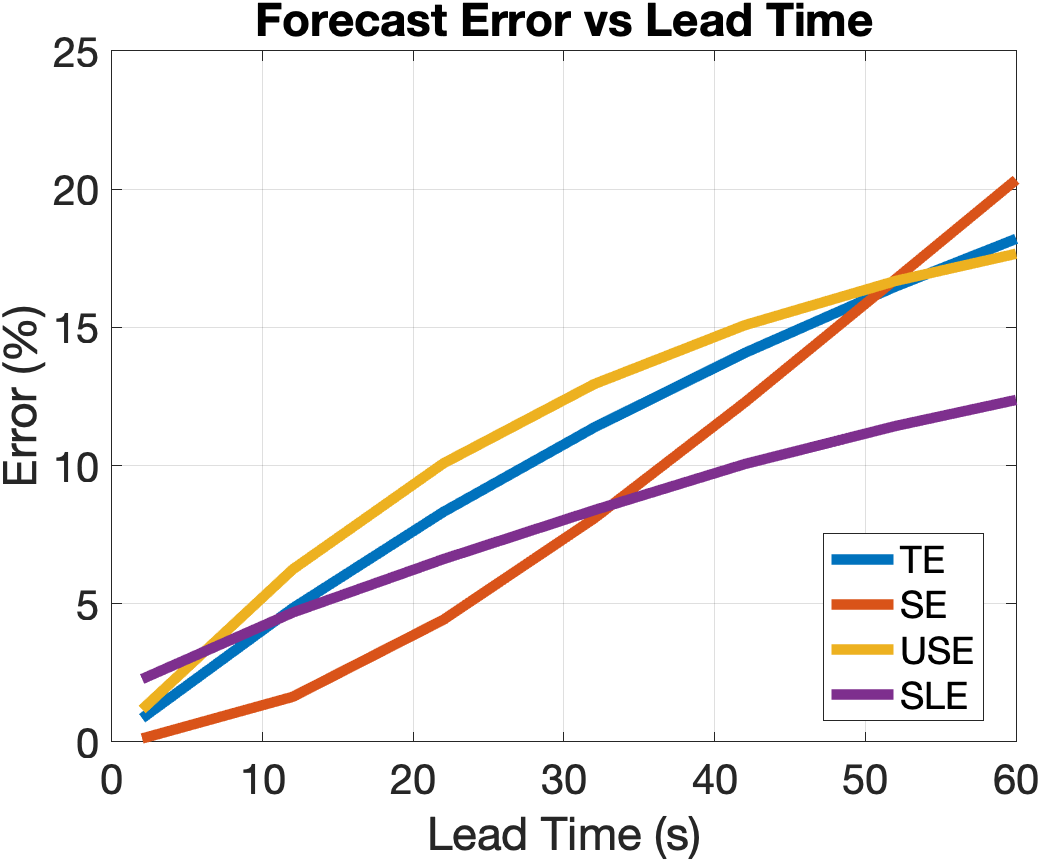}\label{fig:armaforacc}}
\hfil
\subfloat[Correlation of Slopes of AGC]{\includegraphics[width=0.7\columnwidth]{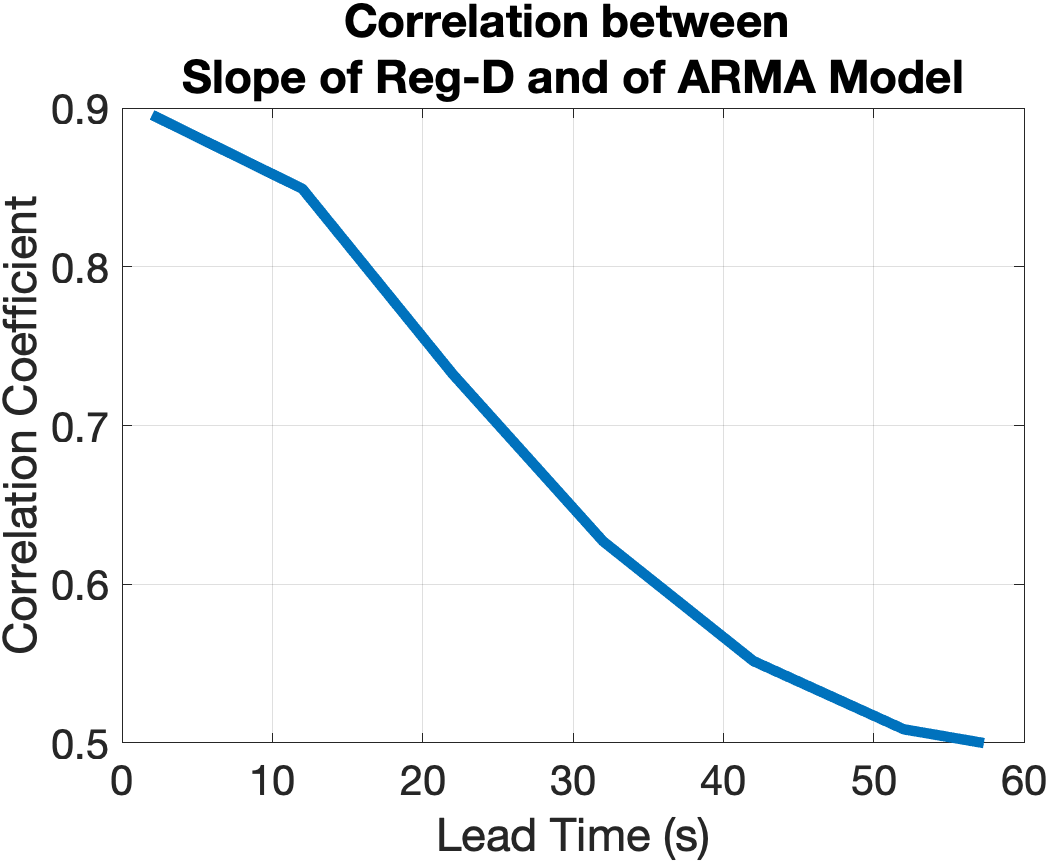} \label{fig:armacorr}}
\caption{Effectiveness of ARMA forecast as a function of lead time}
\end{figure}

The effectiveness of the ARMA model depends highly on the lead time of the forecast. To quantify this, the ARMA model was subjected to varying lead times, and the accuracy of the forecasts was evaluated for twelve one-hour Reg-D signals from all the months of the year, \textcolor{black}{sampled at 2-s intervals.  That is, the ARMA model was tested on $12\times 60\times 30=21600$ samples of the Reg-D signal. } The mean results are shown in Fig. \ref{fig:armaforacc}, where TE stands for the total mean absolute error (MAE) between the output of the ARMA model and that of the Reg-D signal, SE stands \textcolor{black}{for the MAE considering only the samples when Reg-D is saturated at -1 or 1, USE stands for MAE considering only the samples when Reg-D is not saturated. SLE stands for slope error, which is the MAE between the slopes of the Reg-D signal and that of the output of the ARMA model, with the slope computed over a specified lead time.} It can be seen that all the errors increase as the lead time is increased. However, the error when the signal is saturated (SE), while lower than other errors, increases sharply with the increase in lead time, while the error between the slopes (SLE) increases at a slower rate when the lead time increases. Hence, when only the direction is required from the AGC signal, the ARMA model can be effective even for a relatively large lead time. From these results, it is found that in any case, the error in predicting the value of Reg-D is less than 15\%, for up to 30 s. \textcolor{black}{Moreover, the ARMA model was tested against common neural network (NN) architectures: dense feedforward NN, recurrent NN (RNN), long short term memory (LSTM), and gated recurrent unit (GRU), each with two layers of 32 units each, and the results were found to be within 1\% of their predictions (Fig. \ref{fig:nn}), indicating that it is a useful and systematic approach to forecasting compared to black box-based data-driven NN models.}

\subsubsection{Correlation of slopes vs. lead time}
Occasionally, it is necessary to only predict the slope of the AGC signal and not its exact value. Such a case may arise when we want to make a decision based on only the future \textit{direction} of the AGC signal. For example, if the AGC signal is predicted to be lower in value in the future, we can use that information to pre-compensate the AGC signal such that the distributed resources are properly utilized. Hence, a study was conducted to find how much the slope of the ARMA model and that of the original AGC signal are correlated.

The effect of lead times on the correlation between the slopes of the output of the ARMA model and Reg-D is shown in Fig. \ref{fig:armacorr} for the same one-hour AGC signals considered above. It can be seen that the mean correlation decreases mostly linearly as the lead time increases to about 30-time steps or 1 min. Hence, the AR(3) model can predict the slope of Reg-D fairly accurately within less than half a minute.

\subsubsection{Detection of Slopes}
\textcolor{black}{The correlation between slopes of AGC as described above can be used to \textit{classify} the AGC signal into three classes: ``Up", ``Down", and ``Flat", based on whether it is moving up, going down, or remaining flat respectively.} To define what is meant by  \textit{flat}, a certain threshold of slope may be accepted, so that if the slope is outside or larger than that threshold, the AGC signal will be either determined to move up or go down in the future. However, if the slope of the AGC signal remains within that threshold band, it will be deemed to remain flat. Of course, the accuracy of the classification would depend on the value of the threshold. Hence, a study was conducted that describes the effect of the threshold on the classification by using confusion matrices \cite{forbes1995classification}. The threshold is defined as a certain percentage of the range of slopes of the AGC signals considered.
\begin{figure}[t]
\centering
\subfloat[10\% of Slope Range]{\includegraphics[width=0.6\columnwidth]{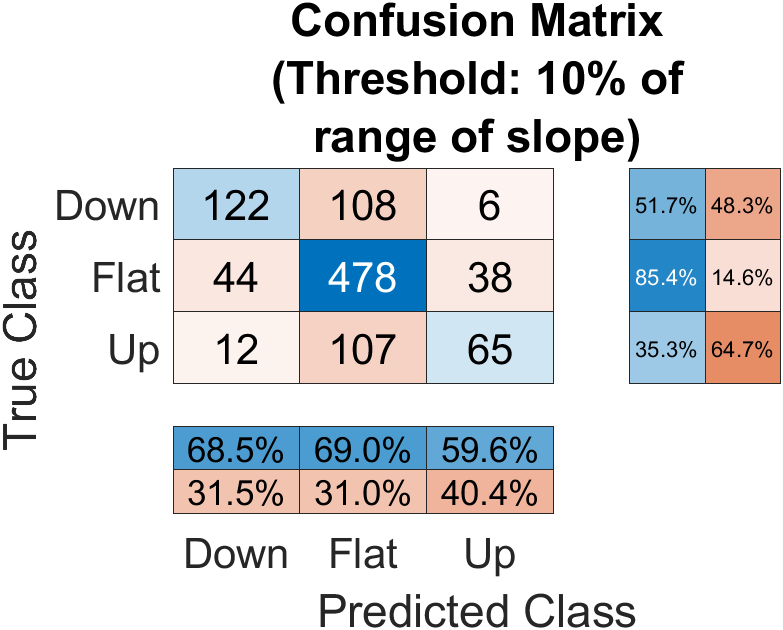}\label{fig:confmat10}}
\hfil
\subfloat[20\% of Slope Range]{\includegraphics[width=0.6\columnwidth]{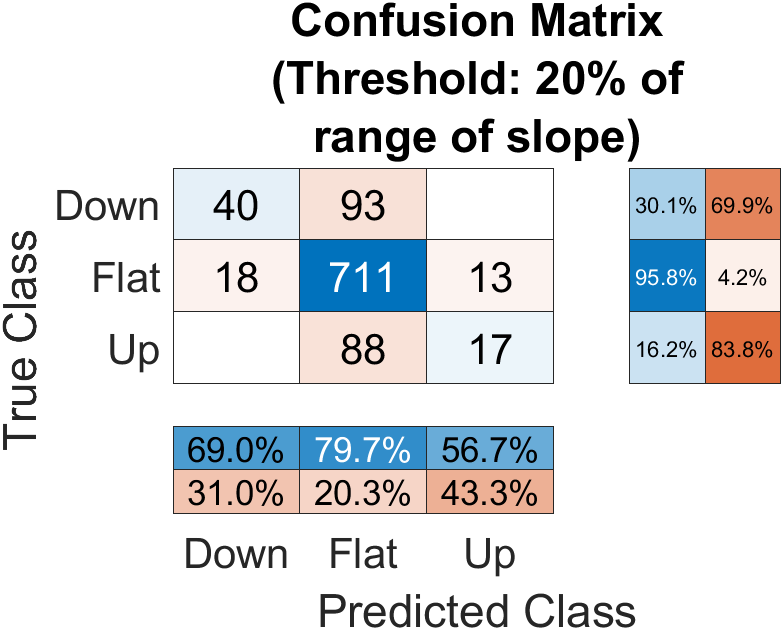}\label{fig:confmat20}}
\caption{Confusion Matrix for Detection of Slope}
\label{confmat}
\end{figure}
The results of one such classification on Reg-D signal are shown in Fig \ref{confmat}. It can be seen that when the threshold is low (10\% of range of slope, as in Fig. \ref{fig:confmat10}), then more cases are wrongly identified as flat, up, or down, but when the threshold is high (20\% of range of slope, as in Fig. \ref{fig:confmat20}), there are fewer errors in classification, although there are more cases which are considered flat. The choice of the proper threshold will be determined by the particular application and the tolerance allowed for the regulating resources.

\subsubsection{Sensitivity of the ARMA model to training set}
To determine if the coefficients of the AR(3) model are sensitive to the training set used for training that ARMA model, different training sets were considered, and the ARMA model was re-fitted on those training sets. Specifically, twelve training sets of Reg-D were considered from July 2018 to June 2019, one for each month of the year. The results are shown in Fig. \ref{fig:armarob}. It can be seen that the three autoregressive coefficients of the AR(3) are relatively flat with respect to the training sets. \textcolor{black}{This provides confidence that the ARMA model is fairly insensitive to the training sets, and thus, to the specific portion of the AGC used for training.}

\subsection{VARMA Modeling}
The ARMA model developed in the previous subsection can be further improved if we have other time-series information. Typically, the power system grid frequency data is available along with the AGC signal. Since the AGC signal is generated to control the frequency, there is a high correlation between the frequency and the AGC signal. This correlation can be utilized to form a \textit{multivariate} time series forecasting model to improve the forecasts that would be otherwise generated by an ARMA model \cite{reinsel2003elements}.

\subsubsection{Cross Correlation of Reg-D with measured grid frequency}
\begin{figure}[t]
\centering
\subfloat[Sensitivity of ARMA coefficients]{\includegraphics[width=0.8\columnwidth]{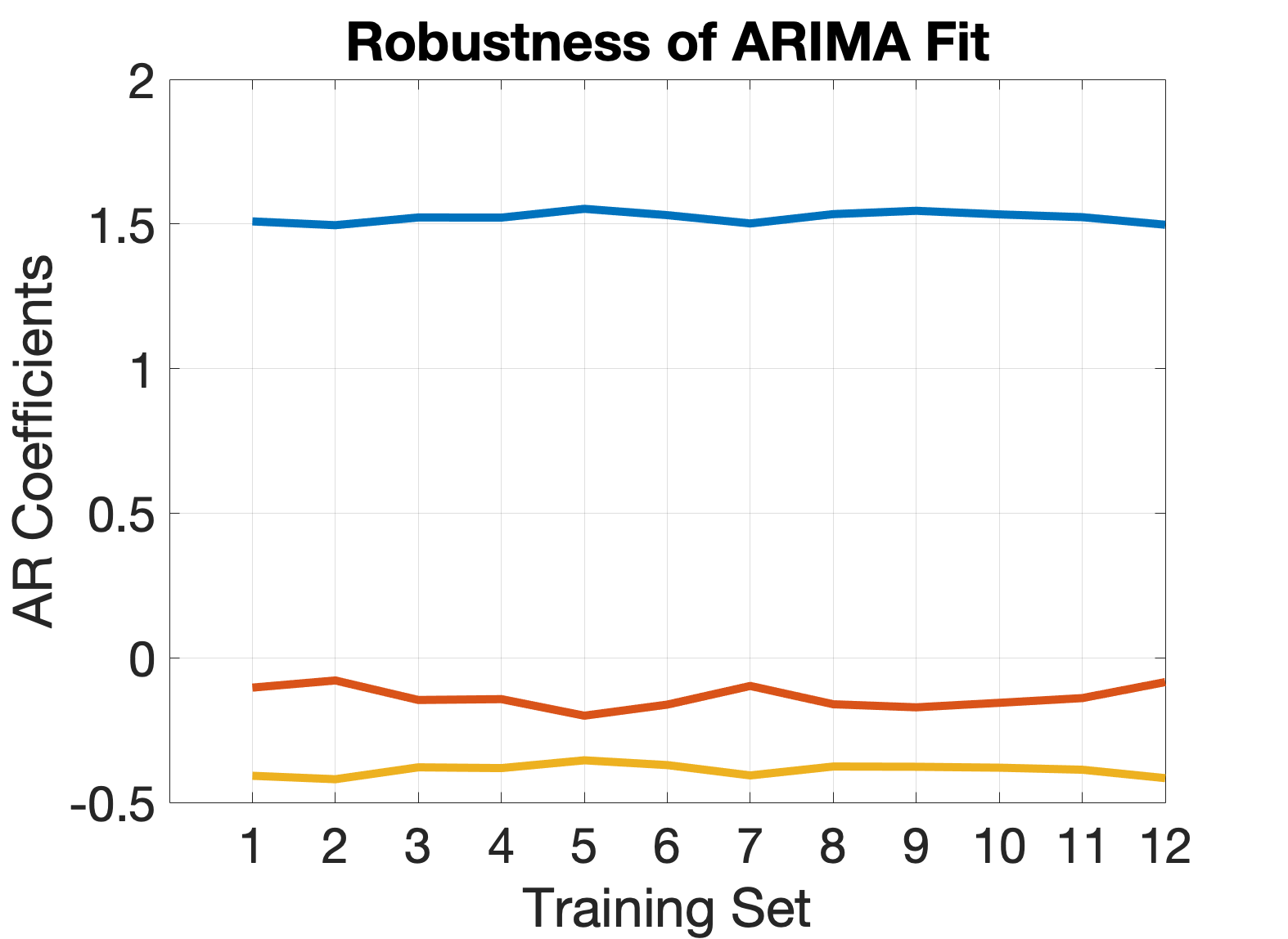}\label{fig:armarob}}
\hfil
\subfloat[Filtering Frequency]{\includegraphics[width=0.7\columnwidth]{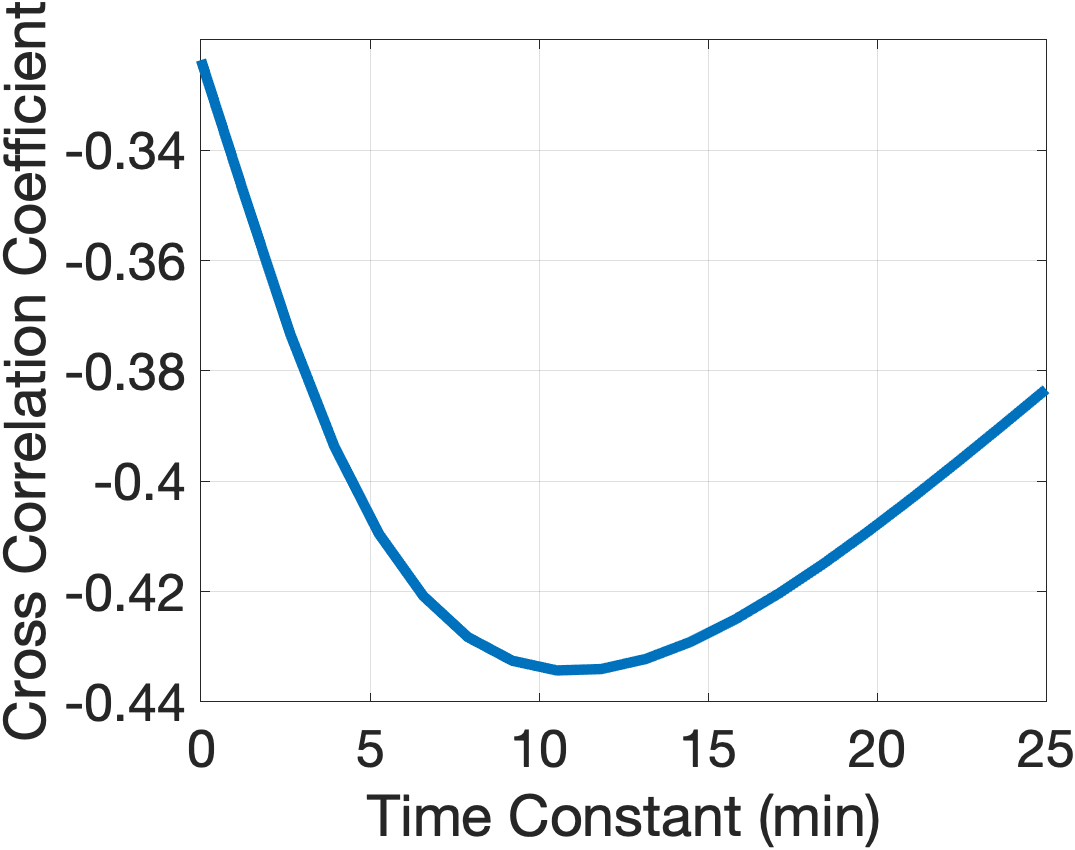}\label{fig:tauagc}}
\caption{Robustness of ARMA model and Filtering Frequency for VARMA model}
\label{agcfreq}
\end{figure}

Grid frequency data were available for one day, 20th June 2019, at a sample time of 100 ms from The University of Tennessee Knoxville. The data was collected on the same grid as Reg-D was used. Since the frequency data were sampled every 100 milliseconds while the AGC signal is sampled every 2 seconds, to enable the highest utilization of information, the frequency data needs to be filtered appropriately. 
A first-order lag filter was chosen for filtering the frequency data. To select the time constant optimally, the cross-correlation between frequency and Reg-D was evaluated, after the frequency was filtered with a first-order transfer function, $1/(\tau s+1)$, with the specified time constant, $\tau$. The result is shown in Fig. \ref{fig:tauagc}. It can be seen that a time constant of 10 min leads to the highest magnitude of cross-correlation between the (filtered) frequency and the AGC signal. Hence, this filtered frequency is used to fit the VARMA model described below. \textcolor{black}{Note that the cross-correlation is negative as expected since when the frequency is \textit{low}, there is not enough generation, thus requiring a \textit{high} value for AGC to balance demand and supply.}

\begin{figure}[t]

    \centering
    \includegraphics[width=1.05\columnwidth]{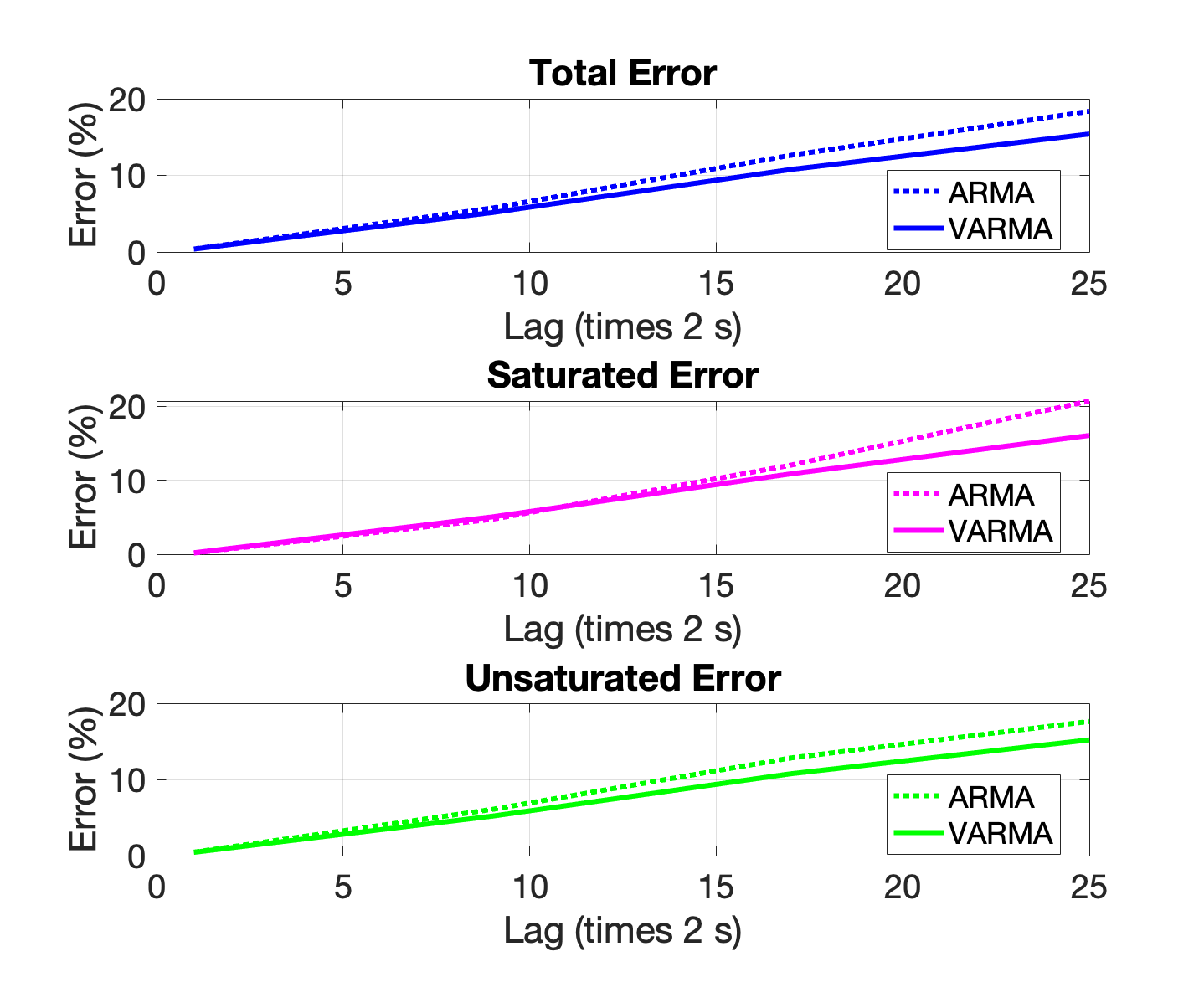}
    \caption{Effectiveness of VAR(3) model}
    \label{fig:var3}
\end{figure}
\subsubsection{Model}
To be consistent with the ARMA model considered earlier, the VARMA model considered here is a VAR($g$) model \cite{lutkepohl2006forecasting}, given by:
\begin{equation*}
    \mathbf{r}[k]	=\boldsymbol{\mu}+\Phi_{1}\mathbf{r}[k-1]+\Phi_{2}\mathbf{r}[k-2]
	+...+\Phi_{g}\mathbf{r}[k-g]+\mathbf{a}\left[k\right]
\end{equation*}
where, $g$ is the number of lags, e.g., 3 as considered for the ARMA model, $\mathbf{r}[k]$ is the multivariate time series (Reg-D and frequency in this case), $\Phi_{i}$ are autoregressive coefficient matrices, and $\mathbf{a}[k]$ is the multivariate Gaussian innovation with covariance matrix $\Sigma$ and $\boldsymbol{\mu}$ is the mean level vector of the process. The coefficients can be obtained using maximum likelihood estimation, for example, using \texttt{estimate} command of MATLAB after creating a VARMA model using \texttt{varm}.

\subsubsection{Effectiveness of VARMA forecast}
To investigate the effectiveness of the VAR(3) model, experiments were conducted on a four-hour Reg-D signal and a corresponding length of the frequency signal on 20th June 2019, to find the total error (TE), saturated error (SE), and unsaturated error (USE), as defined in Section \ref{ssec:forarmaacc}, as a function of the lead times. It can be seen from Fig. \ref{fig:var3} that VAR(3) provides 3.5\% less TE, 5\% less SE, and about  2\% less USE than the corresponding AR(3) model. This shows that VAR(3) is more accurate in predicting the AGC signal compared to the corresponding AR(3) model.

\subsubsection{Application of VARMA model to Model Predictive Control Framework}
\begin{figure}[t]
    \centering
    \includegraphics[width=\columnwidth]{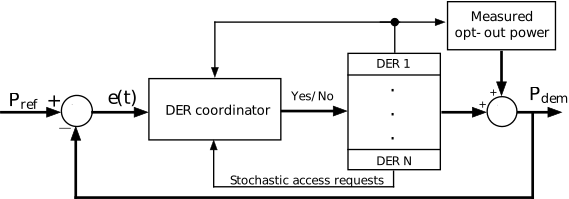}
    \caption{\textcolor{black}{Closed-loop feedback system for PEM with the reference power $P_{\rm{ref}}$ provided by the grid or market operator and the aggregate net-load $P_{\rm{dem}}$ measured by the coordinator.}}
    \label{fig:pem_schem}
\end{figure}

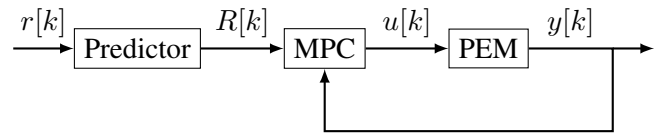
\begin{figure}[!t]
\centering
\begin{tikzpicture}[>=latex,scale=1.1, transform shape]
\node[draw] (mpc) at (0,0){MPC};
\node[draw] (pem) [right= of mpc]{PEM};
\node[draw] (pred) [left= of mpc]{Predictor};
\draw[->,thick] (pred)--node[above]{$R[k]$}(mpc);
\draw[->,thick] (mpc)--node[above]{$u[k]$}(pem);
\draw[->,thick] ($(pred)-(1.5,0)$)--node[above]{$r[k]$}(pred);
\draw[->,thick] (pem)--node[above]{$y[k]$}++(1.5,0)--++(0,-1)-|(mpc);
\draw[->,thick] ($(pem)+(1.5,0)$)--++(0.5,0);
\end{tikzpicture}
\caption{\textcolor{black}{MPC-based Precompensator}}
\label{fig:mpc}
\end{figure}

\textcolor{black}{The VARMA forecasting model can be applied in a predictive control setting, where it can provide valuable forecasts of the future AGC signal to generate optimal control decisions for the regulating grid resources, especially when their outputs are down/up-ramp limited. For example, in \cite{brahma2021optimal}}, a model predictive controller (MPC) was designed by the authors to ensure optimal tracking performance of the Packetized Energy Management (PEM) scheme, and its performance was tested using ARMA forecasts. \textcolor{black}{PEM (Fig. \ref{fig:pem_schem}) is a demand dispatch scheme that can be used to provide ancillary services such as frequency regulation. In PEM, DERs stochastically request access for power from a DER coordinator, which then grants or rejects them uninterruptible access to the grid for a specified period called a packet length. Details on PEM can be found in \cite{Almassalkhi:2018IMA}. A characteristic of PEM is that once packet requests are accepted by the coordinator, it locks devices ON for the duration of their packet length. This causes the aggregate response of DERs to become down ramp-limited, and consequently low tracking performance while tracking down ramps in the AGC signal. The MPC design (Fig. \ref{fig:mpc}) in \cite{brahma2021optimal} overcomes this issue and improves the tracking performance of PEM while ensuring less device switching. The objective of this subsection is to investigate whether the tracking performance of PEM with MPC can be improved using VARMA forecasts compared to ARMA forecasts presented in \cite{brahma2021optimal} (which the reader is encouraged to refer to for background and context).} 

\begin{figure}[t]
    \centering
    \includegraphics[width=0.8\columnwidth]{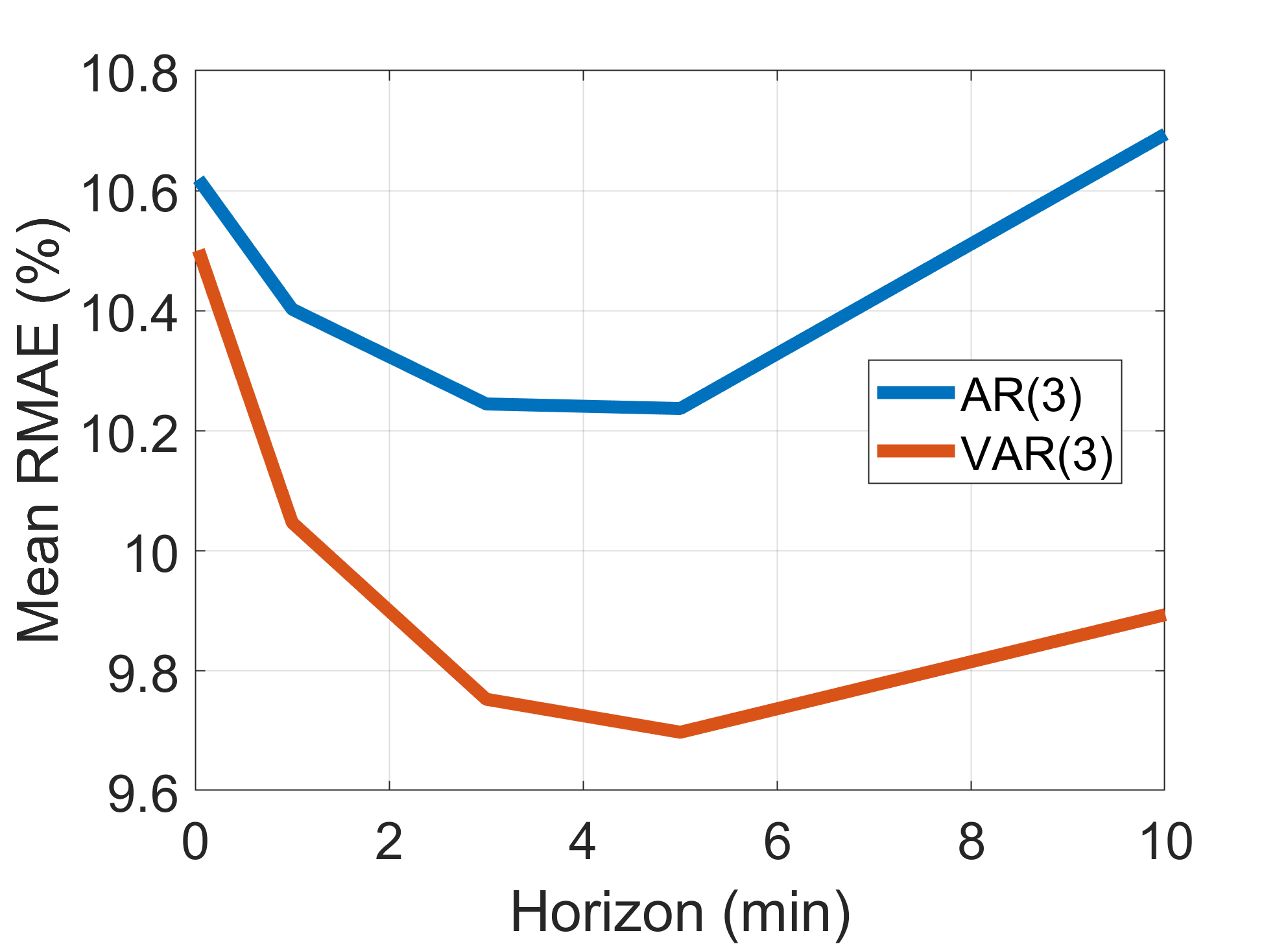}
    \caption{Effectiveness of MPC with VARMA forecast}
    \label{fig:varm}
\end{figure}

To investigate the effect of improved forecasts from the VARMA model on the tracking performance of PEM, simulations were conducted on PEM, equipped with the MPC but this time with VAR(3) forecasts using \textit{both} Reg-D and frequency data. Four representative 1-h datasets of Reg-D and frequency on 20th June 2019 were chosen (specifically, 6-7 AM, 12-1 PM, 6-7 PM, and 12-1 AM EST) for the simulations, and the MPC horizon was varied. The average relative mean absolute tracking errors (RMAE) are shown in Fig. \ref{fig:varm}. It can be seen that RMAE with VAR(3) forecast is smaller by about 0.7\%. With a horizon of 10 min, it can be seen that while the ARMA forecast performs worse than with no MPC (horizon tending to 0), the VAR(3) forecast improves and also results in RMAE lesser by 0.8\% than the corresponding AR(3) forecast. This indicates that utilizing additional information from the frequency data results in improved forecasts even at a high prediction horizon, which then leads to improved tracking with the MPC.

\section{Conclusion}
In this paper, a statistical model and a time series-based forecasting model are developed for the modeling and forecasting of AGC signals \textcolor{black}{to provide a useful starting point for designing model-based controllers for frequency regulating units.} By conducting a statistical analysis on a widely used AGC regulation signal, PJM Reg-D, including its variability, power spectrum, and saturation, a stochastic model, driven by stationary white noise, is derived that is shown to fairly accurately model the Reg-D signal and capture its second moments and saturated nature. By conducting studies on the autocorrelation and partial autocorrelation functions, an ARMA model is derived that fairly accurately forecasts the Reg-D signal within less than half a minute, both directionally and in predicting its value. Further, by incorporating information from the power grid frequency data, it can be seen that a VARMA model can further improve the forecasts obtained using AGC data alone. The VARMA forecasts have been used in an MPC framework to improve the tracking performance of a DER coordination scheme compared to ARMA forecasts.

\textcolor{black}{Future work includes investigation of better forecasts and models, including machine learning-based approaches and architectures, studies on different lead times and orders of ARMA and VARMA models, correlation with power load data, testing different probability distributions to model AGC signals from other ISOs, and extending analyses to incorporating more frequency data and locations.}


\section*{Acknowledgment}
\textcolor{black}{The authors would like to acknowledge the support of the U.S. Department of Energy through its  Advanced Research Projects Agency-Energy (ARPA-E) award: DE-AR0000694. The authors would like to thank Dr. Weikang Wang and Prof. Yilu Liu at the University of Tennessee Knoxville (FNET) for providing us with PMU data, as well as Danielle Croop and Anthony Giacomi at PJM for helpful discussions on PJM Performance Scoring.}



%

\bibliographystyle{IEEEtran}
\bibliography{ref}

\end{document}